\documentclass[prd,twocolumn,showpacs,showkeys,preprintnumbers,floatfix,nofootinbib,superscriptaddress]{revtex4-1}
\usepackage{amsfonts} 
\usepackage{amssymb} 
\usepackage{amsmath} 
\usepackage{graphicx} 
\usepackage{subfigure} 
\usepackage{array} 
\usepackage{dcolumn} 
\usepackage{bm} 
\usepackage{latexsym} 
\usepackage{longtable} 
\usepackage{hyperref} 
\usepackage{bbold}
\usepackage{color}
\usepackage{multirow}
\usepackage{rotating}
\usepackage{comment}

\graphicspath{{./figs/}}

\newcommand{\tsepi}{\mathop{\tau \to \infty}\nolimits}

\definecolor{green}{rgb}{0.1, 0.8, 0.1}

\begin{document}


\title{Quark contribution to the proton spin from 2+1+1-flavor lattice QCD}
\author{Huey-Wen~Lin}
\email{hwlin@pa.msu.edu}
\affiliation{Department of Physics and Astronomy, Michigan State University, East Lansing, MI, 48824, USA}
\affiliation{Department of Computational Mathematics,  Science and Engineering, Michigan State University, East Lansing, MI 48824, USA}

\author{Rajan~Gupta}
\email{rajan@lanl.gov}
\affiliation{Los Alamos National Laboratory, Theoretical Division T-2, Los Alamos, NM 87545, USA}

\author{Boram~Yoon}
\email{boram@lanl.gov}
\affiliation{Los Alamos National Laboratory, Theoretical Division T-2, Los Alamos, NM 87545, USA}

\author{Yong-Chull~Jang}
\email{ypj@bnl.gov}
\affiliation{Brookhaven National Laboratory, Physics Department, Upton, NY 87545, USA}

\author{Tanmoy~Bhattacharya}
\email{tanmoy@lanl.gov}
\affiliation{Los Alamos National Laboratory, Theoretical Division T-2, Los Alamos, NM 87545, USA}

\collaboration{PNDME Collaboration}
\preprint{LA-UR-18-25337}
\preprint{MSUHEP-18-010}
\date{\today}
\begin{abstract}
  We present the first chiral-continuum extrapolated up, down and
  strange quark spin contribution to the proton spin using lattice
  QCD.  For the connected contributions, we use eleven ensembles of
  2+1+1-flavor of Highly Improved Staggered Quarks (HISQ) generated
  by the MILC Collaboration. They cover four lattice spacings $a \approx
  \{0.15,0.12,0.09,0.06\}$~fm and three pion masses, $M_\pi \approx
  \{315,220,135\}$~MeV, of which two are at the physical pion mass.
  The disconnected strange calculations are done on seven of these
  ensembles, covering the four lattice spacings but only one with the
  physical pion mass. The disconnected light quark calculation was
  done on six ensembles at two values of $M_\pi \approx  \{315,220\}$~MeV. 
  High-statistics estimates on each ensemble for all
  three quantities allow us to quantify systematic uncertainties and
  perform a simultaneous chiral-continuum extrapolation in the lattice
  spacing and the light-quark mass.  Our final results are $\Delta u
  \equiv \langle 1 \rangle_{\Delta u^+} = 0.777(25)(30)$, $\Delta d
  \equiv \langle 1 \rangle_{\Delta d^+} = -0.438(18)(30)$, and $\Delta
  s \equiv \langle 1 \rangle_{\Delta s^+} = -0.053(8)$, adding up to a
  total quark contribution to proton spin of $\sum_{q=u,d,s}
  (\frac{1}{2} \Delta q) = 0.143(31)(36)$. The second error is the
  systematic uncertainty associated with the chiral-continuum
  extrapolation. These results are obtained without model assumptions and are
  in good agreement with the recent COMPASS analysis $0.13 <
  \frac{1}{2} \Delta \Sigma < 0.18$, and with the $\Delta q$ obtained
  from various global analyses of polarized beam or target data.
\end{abstract}
\maketitle
%
%
%
%
\section{Introduction}
\label{sec:into}

In 1987, the European Muon Collaboration measured the spin asymmetry
in polarized deep inelastic scattering and presented the remarkable
result that the sum of the spins of the quarks contributes less than
half of the total spin of the proton~\cite{Ashman:1987hv}. This
unexpected result was termed the ``proton spin crisis''.  Lattice QCD
can unravel the mystery of where the proton gets its spin by measuring
the matrix elements of appropriate quark and gluon operators within
the nucleon state. In this paper, we present the first lattice
calculation of the contribution of the intrinsic spin of the quarks to
the proton spin with high-statistics and control over systematic
errors. Our result, $\sum_{q=u,d,s} \frac{1}{2} {\Delta q} =
0.143(31)(36)$, is in good agreement with the COMPASS analysis $0.13 <
\frac{1}{2} \Delta \Sigma < 0.18$ at
3~GeV${}^2$~\cite{Adolph:2015saz}. Note that, above 3~GeV${}^2$ the
change of the axial charges with scale is negligible.

To calculate the nucleon spin using lattice
QCD, one starts with Ji's sum rule~\cite{Ji:1996ek} that provides a
gauge invariant decomposition of the nucleon's total spin as
\begin{equation}
\frac{1}{2} =  \sum_{q=u,d,s,c,\cdot} \left(\frac{1}{2} {\Delta q} + L_q \right) + J_g 
\label{eq:Ji}
\end{equation}
where ${\Delta q} \equiv  {\Delta \Sigma_q} \equiv \langle 1 \rangle_{\Delta q^+} \equiv g_A^q$ 
is the contribution of the intrinsic
spin of a quark with flavor $q$; $L_q$ is the orbital angular momentum
of that quark; and $J_g$ is the total angular momentum of the gluons.
Thus, to explain the spin of the proton starting from QCD, one needs
to calculate the contributions of all three terms. In this paper we
present results for the relatively better determined first term, 
$\frac{1}{2} \Delta \Sigma \equiv \sum_{q=u,d,s} \frac{1}{2} {\Delta q} $. 
 \looseness-1

On the lattice, the axial charge $g_A^q$ is given by 
the matrix element of the flavor diagonal axial current, $\overline{q} \gamma_\mu
\gamma_5 q$, 
\begin{align}
g_A^q \overline{u}_N \gamma_\mu \gamma_5 u_N \!=\!  \langle N| Z_A \overline{q} \gamma_\mu \gamma_5 q | N \rangle 
\label{eq:gAdef}
\end{align}
where $Z_A$ is the renormalization constant and $u_N$ is the neutron spinor. 
In addition to quantifying the contribution of the quarks to the nucleon
spin, 
\begin{equation}
g_A^q \equiv \Delta q \!=\! \int_0^1 dx (\Delta q(x) + \Delta \overline{q} (x) )
\end{equation}
is also the first Mellin moment of the polarized
parton distribution function (PDF) integrated over the momentum
fraction $x$~\cite{Lin:2017snn}.  The charges, $g_A^{u,d,s}$, also
quantify the strength of the spin-dependent interaction of dark matter
with nucleons~\cite{Fitzpatrick:2012ib,Hill:2014yxa}.  Of these,
$\Delta s$ is the least well known and current
analyses~\cite{Lin:2017snn} often rely on assumptions such as SU(3)
symmetry and $\Delta s = \Delta \overline{s}$.

\section{Lattice Methodology}
\label{sec:Methodology}

\begin{figure}[tb]
  \subfigure{
    \includegraphics[height=0.8in]{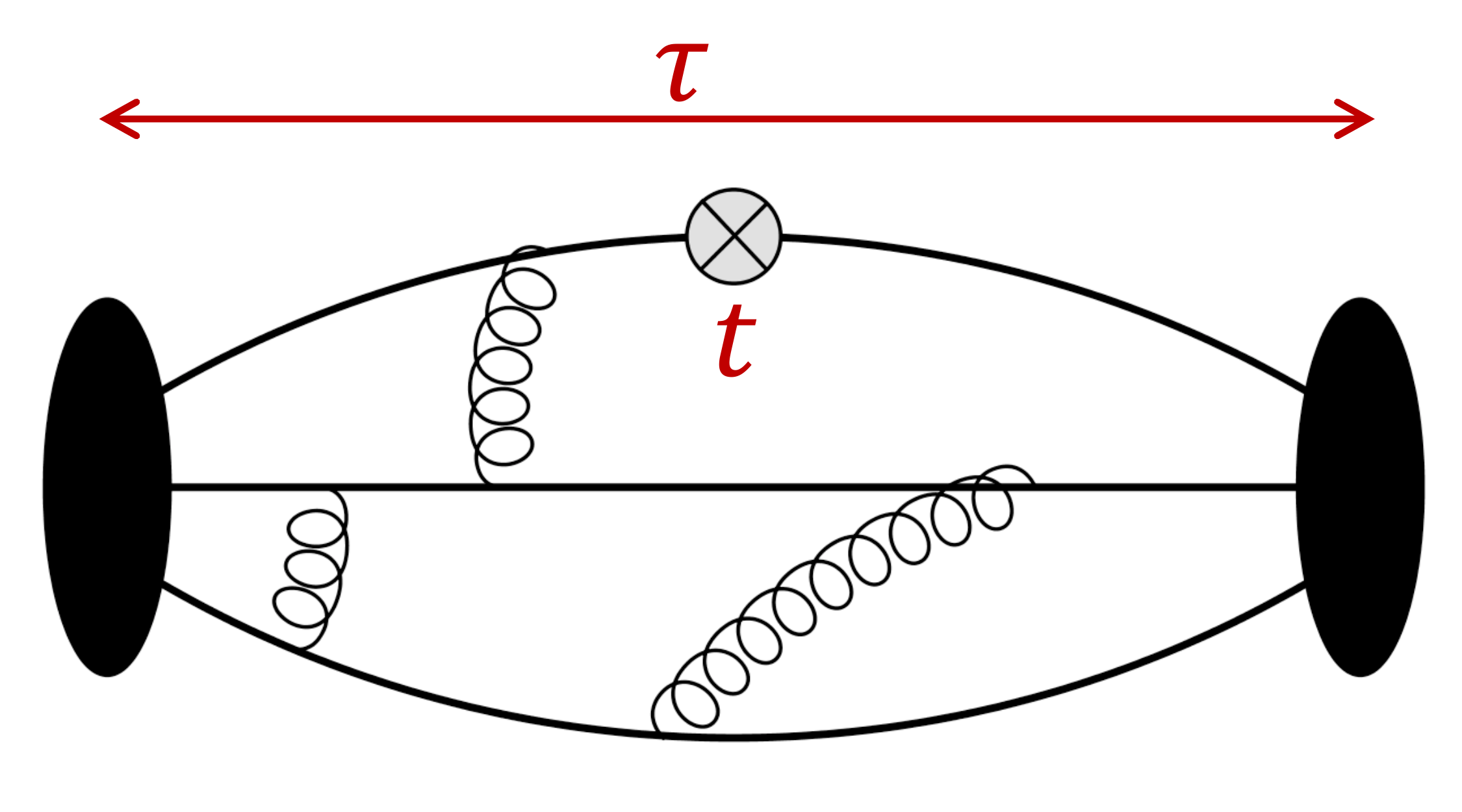}
 \hspace{0.1\linewidth}
    \includegraphics[height=1.0in]{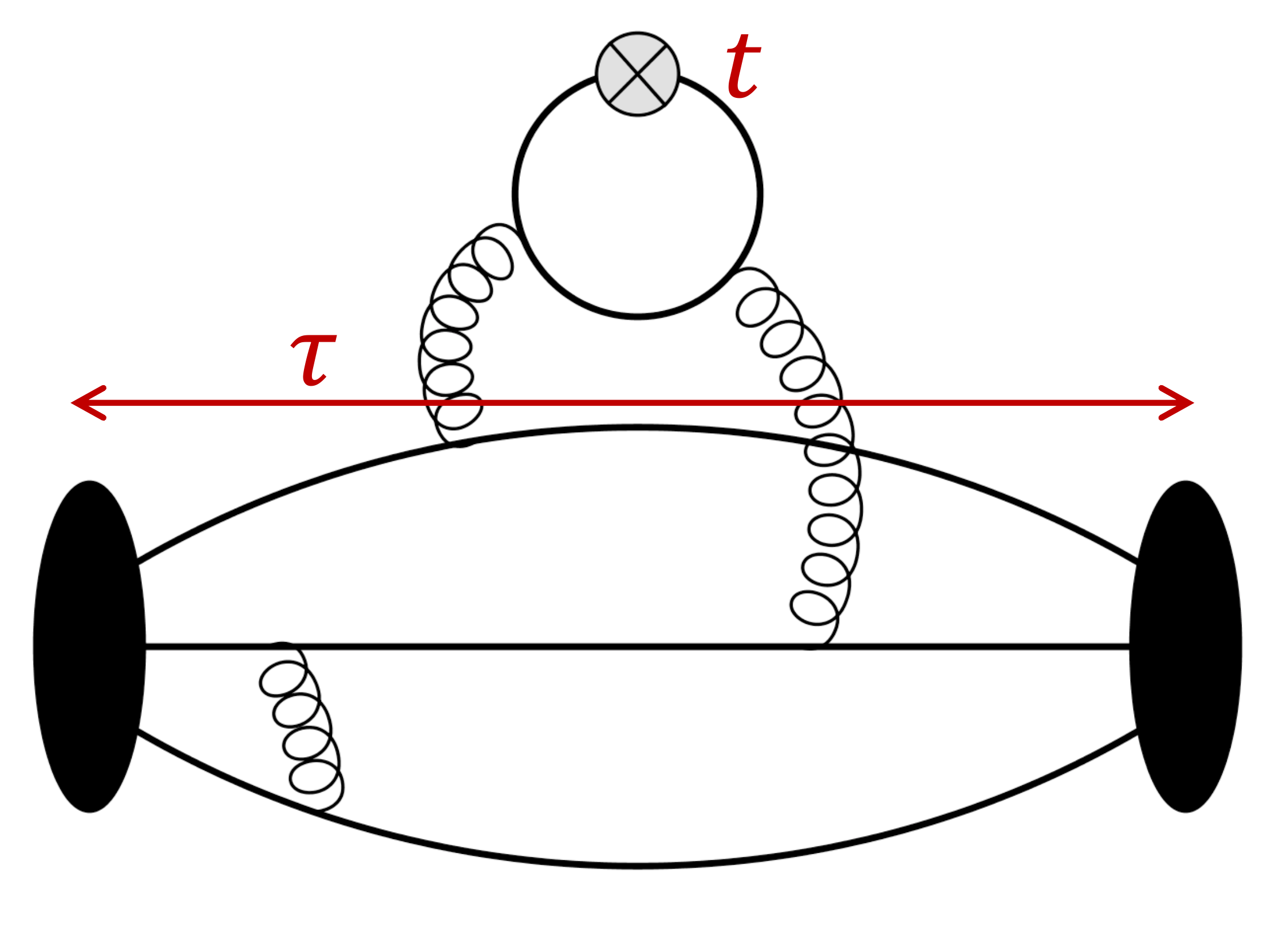}
  }
\vspace{-0.1in}
\caption{The connected (left) and disconnected (right) three-point
  diagrams that contribute to the flavor diagonal matrix elements of
  the axial operator (labeled by $\otimes$ at time slice $t$) within
  the nucleon state. The black blobs denote nucleon source and sink, 
  separated by Euclidean time $\tau$.\looseness-1
  \label{fig:con_disc}}
\vspace{-0.1in}
\end{figure}

The calculation of the flavor diagonal charges $g_A^q$ is now
mature~\cite{Bhattacharya:2015wna,Gupta:2018qil}. The challenge is to
obtain high-statistics results for both the connected and disconnected
contributions to nucleon three-point functions illustrated in
Fig.~\ref{fig:con_disc} and then address the various systematics. An
important finding of this work is that the lattice discretization
errors and the chiral corrections are large; consequently the
evaluation of the renormalized charges at the physical pion mass,
$M_{\pi^0} = 135$~MeV, and the extrapolation to the continuum limit are 
essential as discussed in Sec.~\ref{sec:CCFV}.\looseness-1

%
%

\begin{table*}[tbp]    
\begin{center}
\renewcommand{\arraystretch}{1.2} 
\begin{ruledtabular}
\begin{tabular}{l|cc|cc|cccc|c}
Ensemble ID     & $a$ (fm)   & $M_\pi$ (MeV) & $L^3\times T$   & $M_\pi L$ & $N_\text{conf}^l$ & $N_\text{src}^l$ & $N_\text{conf}^s$ & $N_\text{src}^s$  &  $N_{\rm LP}/N_{\rm HP}$  \\
\hline
$a15m310 $      & 0.1510(20) & 320(5)        & $16^3\times 48$ & 3.93      & 1917              & 2000             & 1919              &  2000             &  50  \\
\hline
$a12m310 $      & 0.1207(11) & 310(3)        & $24^3\times 64$ & 4.55      & 1013              & 5000             & 1013              &  1500             &  30   \\
$a12m220 $      & 0.1184(10) & 228(2)        & $32^3\times 64$ & 4.38      &  958              & 11000            & 958               &  4000             &  30   \\
\hline                                                                                                          
$a09m310 $      & 0.0888(08) & 313(3)        & $32^3\times 96$ & 4.51      & 1081              & 4000             & 1081              &  2000             &  30   \\
$a09m220 $      & 0.0872(07) & 226(2)        & $48^3\times 96$ & 4.79      & 712               & 8000             & 847               &  10000            &  30/50   \\
$a09m130 $      & 0.0871(06) & 138(1)        & $64^3\times 96$ & 3.90      &                   &                  & 877               &  10000            &  50   \\
\hline                                                                                                          
$a06m310 $      & 0.0582(04) & 320(2)        & $48^3\times 144$& 4.52      & 830               & 4000             & 200+340           &  5000+10000       &  50   \\
\end{tabular}
\end{ruledtabular}
\caption{Lattice parameters of the seven ensembles analyzed for the
  disconnected contributions. This table gives the number of
  configurations analyzed for the light ($N_\text{conf}^l$) and
  strange ($N_\text{conf}^s$) quarks, the number of random sources
  ($N_\text{src}$) and the ratio $N_{\rm LP}/N_{\rm HP}$ of LP to HP
  solves used to estimate the quark loop on each configuration.  The
  parameters of the 11 ensembles used for the connected contribution
  are given in Table~1 in Ref.~\cite{Gupta:2018qil}.  }
\label{tab:ens}
\end{center}
\end{table*}

%
%
\begin{table*}[htbp]   
\centering
\begin{ruledtabular}
\begin{tabular}{c|cc|cc|cc}
ID      &  $g_A^{l,{\rm bare}}$      & $g_A^{s,{\rm bare}}$       & $g_A^{l}|_{R1}$ & $g_A^{s}|_{R1}$   & $g_A^{l}|_{R2}$ & $g_A^{s}|_{R2}$  \\ 
\hline                                                                                                     
a15m310 &  $-$0.045(4)[0.9] & $-$0.024(2)[1.2]  & $-$0.044(4) &  $-$0.023(2)  & $-$0.045(4) &  $-$0.024(2) \\
\hline                                                                                                     
a12m310 &  $-$0.053(5)[1.2] & $-$0.027(3)[1.1]  & $-$0.051(5) &  $-$0.025(3)  & $-$0.052(4) &  $-$0.026(3) \\
a12m220 &  $-$0.079(9)[0.8] & $-$0.039(6)[0.7]  & $-$0.075(9) &  $-$0.037(6)  & $-$0.077(9) &  $-$0.038(6) \\
\hline                                                                                                     
a09m310 &  $-$0.056(6)[0.8] & $-$0.033(5)[0.9]  & $-$0.053(6) &  $-$0.031(5)  & $-$0.056(6) &  $-$0.033(5) \\
a09m220 &  $-$0.086(9)[1.3] & $-$0.040(6)[1.3]  & $-$0.082(9) &  $-$0.038(6)  & $-$0.085(9) &  $-$0.039(6) \\
a09m130 &                   & $-$0.048(28)[1.3] &             &  $-$0.046(27) &             &  $-$0.047(27)\\
\hline                                                                                                     
a06m310 &  $-$0.068(9)[0.8] & $-$0.027(10)[1.3] & $-$0.066(9) &  $-$0.026(10) & $-$0.068(9) &  $-$0.027(10)\\
\hline                                                                                                     
Extrapolated &              &                   &  $-$0.115(13) [0.28] & $-$0.052(8) [0.17] & $-$0.120(14) [0.20]& $-$0.054(8) [0.21] \\
\end{tabular}
\end{ruledtabular}
\caption{The bare and renormalized charges from the different
  ensembles are given along with the values after extrapolation to
  $a=0$ and $M_\pi=135$~MeV.  The charges, renormalized at 2~GeV in
  the $\overline{MS}$ scheme in the two ways defined in
  Eq.~\protect\eqref{eq:renorm}, are given in columns 4--7.  In all
  cases, the numbers within the square brackets are the $\chi^2/DOF$
  of the fits. In the ESC fits for extracting the bare charges, shown
  in Fig.~\protect\ref{fig:gAdisc}, the $\chi^2/DOF$ with $DOF\approx
  20$ is given in columns 2--3. In the chiral-continuum fits, using
  Eq.~\protect\eqref{eq:extrapgAST} and shown in
  Fig.~\protect\ref{fig:gls-extrap}, the $\chi^2/DOF$ with $DOF=3$
  (light) or 4 (strange) is given in the last row. \looseness-1}
\label{tab:results}
\end{table*}


The calculations of the connected and disconnected contributions to
$g_A^{u,d}$ were done separately using 2+1+1-flavor ensembles of
HISQ fermions~\cite{Follana:2006rc} generated by the MILC
Collaboration~\cite{Bazavov:2012xda}.  The construction of the two- and
three-point correlation functions used in the analysis was carried out
using Wilson-clover fermions. We refer to this as the clover-on-HISQ
lattice formulation, which in the continuum limit is expected to give
results for QCD.  All results presented here are for degenerate $u$ and $d$
quarks, with the $s$ and $c$ quark masses tuned to their physical
values.\looseness-1

Results for the connected contributions have been obtained using
eleven HISQ ensembles that cover the range $0.06 \lesssim a \lesssim
0.15$~fm in the lattice spacing, $135 \lesssim M_\pi \lesssim 320$~MeV
in the pion mass and $3.3 \lesssim M_\pi L \lesssim 5.5$ in the spatial
lattice size expressed in terms of $M_\pi L$. The analysis of the
connected contributions, including the simultaneous
chiral-continuum-finite-volume (CCFV) fits has been presented in
Ref.~\cite{Gupta:2018qil} and the final results are
\begin{align}
g_A^{u-d} &=  \phantom{-}1.218(25)(30)  \,, \nonumber \\
g_A^{u}|_{\rm conn}   &=  \phantom{-}0.895(21)   \,, \nonumber \\
g_A^{d}|_{\rm conn}   &=  -0.320(12) \,.
\label{eq:gAconn}
\end{align}
The second error in $g_A^{u-d}$ represents an estimate of the uncertainty due to 
using the leading order corrections in the CCFV fit ansatz. \looseness-1

The computationally expensive calculations of the disconnected
contributions has been carried out on six (for light $u$ and $d$
quark contributions) and seven (for strange quark) HISQ ensembles, as described in
Table~\ref{tab:ens}.  The calculation of the vacuum polarization loop
with the current insertion in the disconnected diagram is carried out
stochastically using Gaussian or $Z_4$ random sources on each
background gauge configuration as described in
Ref.~\cite{Bhattacharya:2015wna}. In this method, the final statistical error is a
combination of the error in the stochastic evaluation on each
configuration and the error due to the average over the gauge
configurations required by the path integral.

To increase the statistics in a cost-effective manner, the calculations of both
the two- and three-point nucleon correlation functions were carried out
using the truncated solver method with bias
correction~\cite{Bali:2009hu,Blum:2012uh}. In this method, correlation
functions are constructed using quark propagators inverted with low
precision (LP) stopping criteria between $r_{\rm LP} \equiv |{\rm
  residue}|_{\rm LP}/|{\rm source}| = 10^{-3}$ and $5 \times 10^{-4}$, and 
high-`precision (HP) with $r_{\rm HP}$ between $10^{-7}$ and
$10^{-8}$~\cite{Bhattacharya:2015wna,Gupta:2018qil}.  The bias
corrected correlation functions on each configuration are given
by\looseness-1
\begin{equation}
 C^\text{imp} 
 = \sum_{i=1}^{N_\text{LP}} 
   \frac{  C_\text{LP}(\mathbf{x}_i^\text{LP})}{N_\text{LP}} 
  + \sum_{i=1}^{N_\text{HP}} \left[
    \frac{C_\text{HP}(\mathbf{x}_i^\text{HP})
    - C_\text{LP}(\mathbf{x}_i^\text{HP})}{N_\text{HP}} 
    \right] \,, \nonumber
  \label{eq:2-3pt_AMA}
\end{equation}
where $C_\text{LP}$ and $C_\text{HP}$ are the two- or three-point
functions calculated in LP and HP, respectively, and
$\mathbf{x}_i^\text{LP}$ and $\mathbf{x}_i^\text{HP}$ are the source
positions for the two kinds of propagator inversion. The bias was
found to be smaller than the statistical errors in all
cases.\looseness-1

\begin{figure*}[hptb]
\begin{center}                                               
  \subfigure{
    \includegraphics[height=1.0in,trim={0.08cm 0.70cm 0 0.06cm},clip]{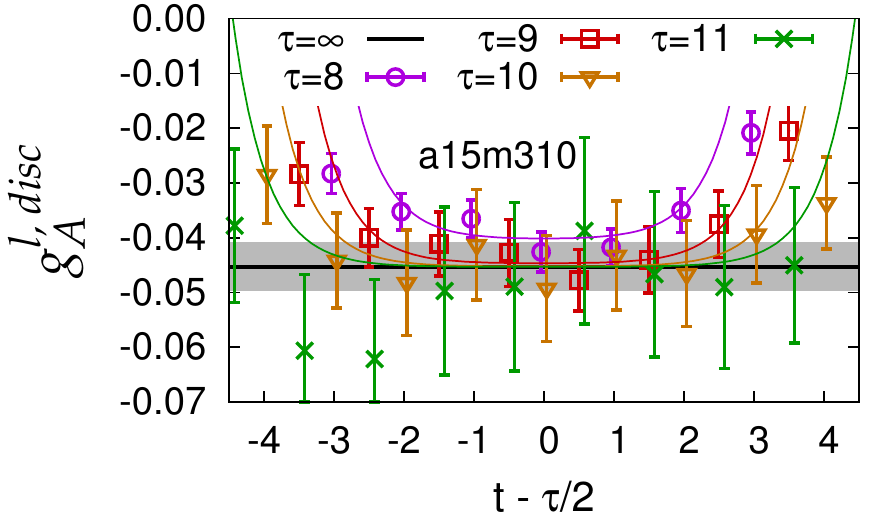}
    \includegraphics[height=1.0in,trim={1.08cm 0.70cm 0 0.06cm},clip]{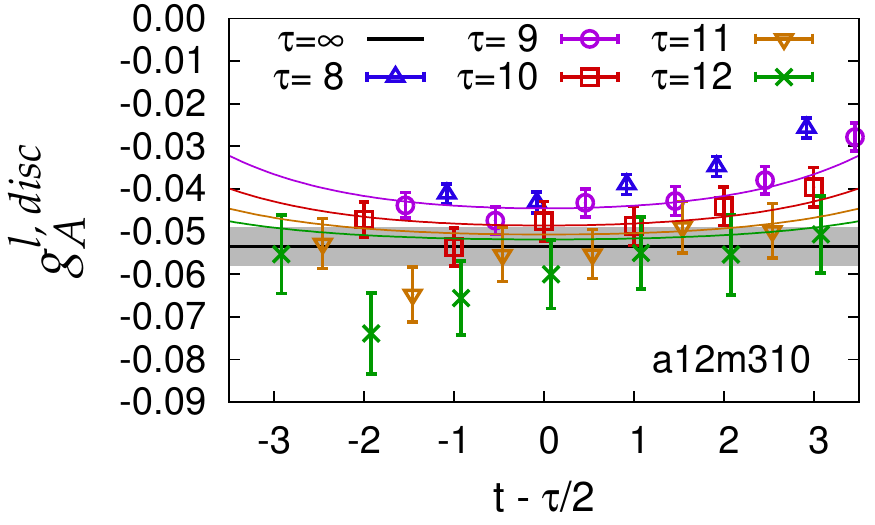}
    \includegraphics[height=1.0in,trim={1.08cm 0.70cm 0 0.06cm},clip]{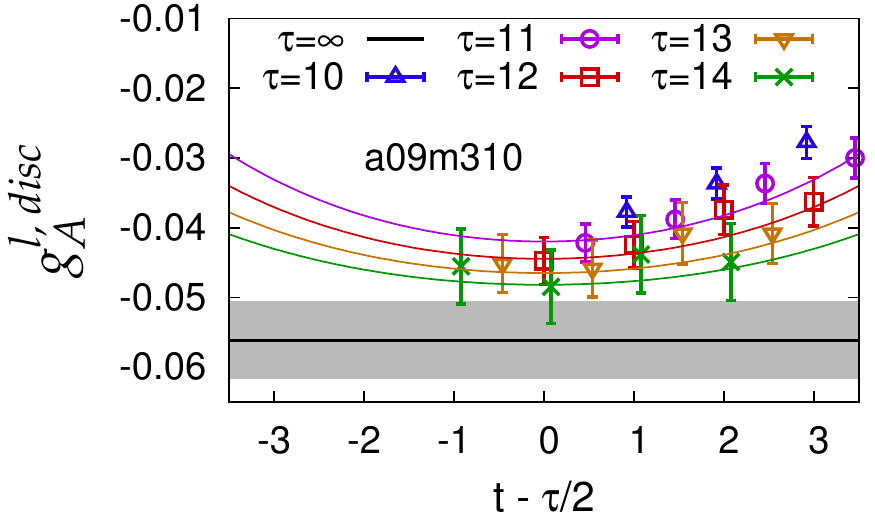}
    \includegraphics[height=1.0in,trim={1.08cm 0.70cm 0 0.06cm},clip]{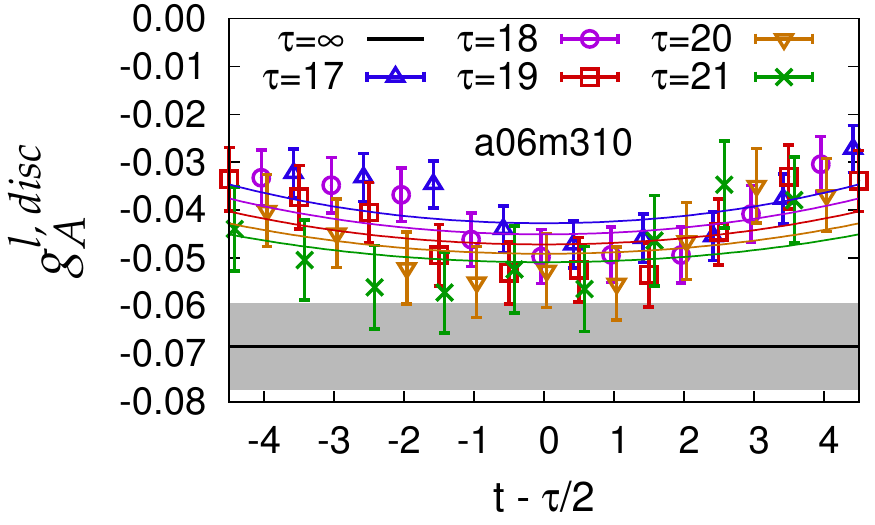}
  }\\
  \subfigure{
    \includegraphics[height=1.2in,trim={0.08cm 0.10cm 0 0.06cm},clip]{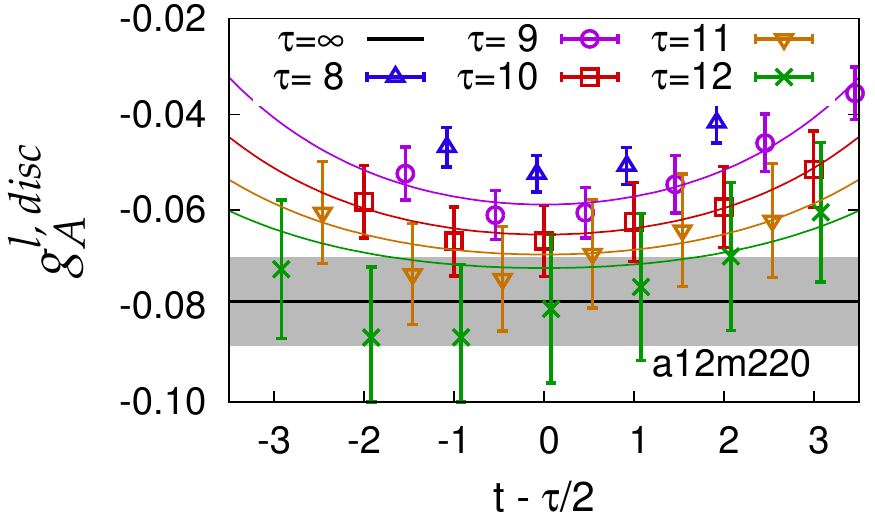}
    \includegraphics[height=1.2in,trim={1.08cm 0.10cm 0 0.06cm},clip]{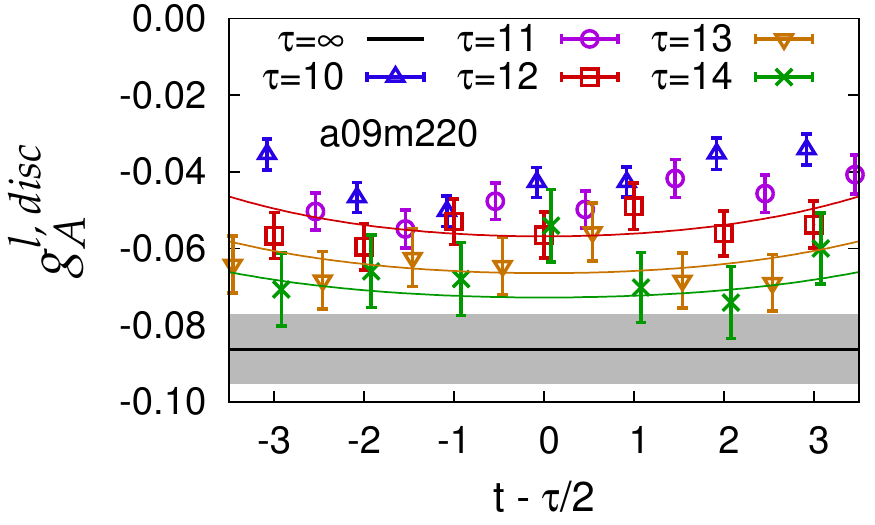}
  }
  \subfigure{
    \includegraphics[height=1.0in,trim={0.08cm 0.70cm 0 0.06cm},clip]{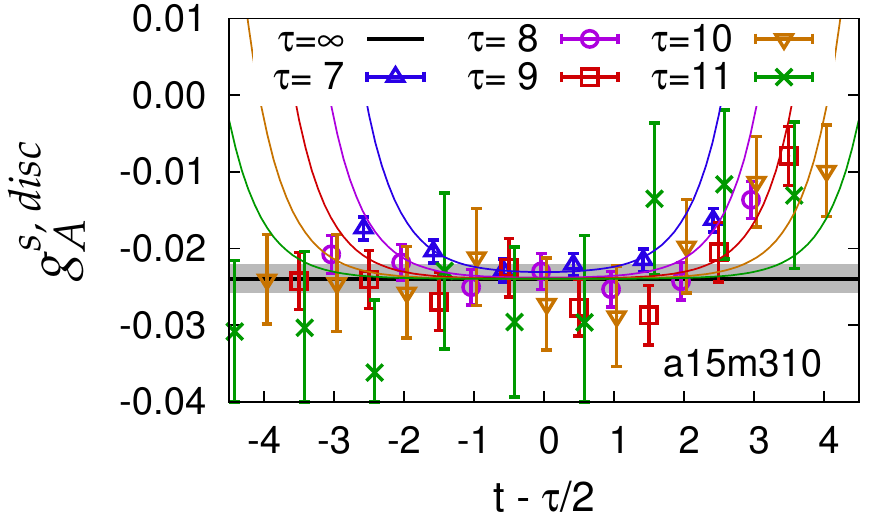}
    \includegraphics[height=1.0in,trim={1.08cm 0.70cm 0 0.06cm},clip]{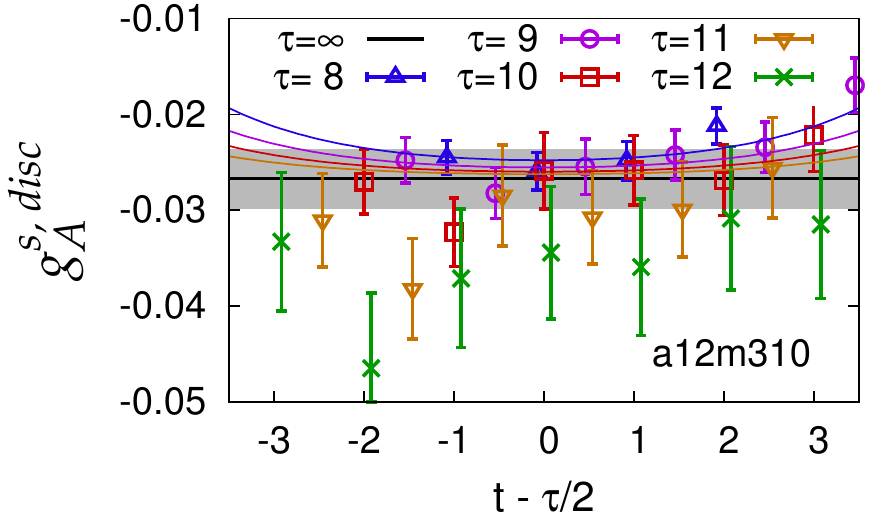}
    \includegraphics[height=1.0in,trim={1.08cm 0.70cm 0 0.06cm},clip]{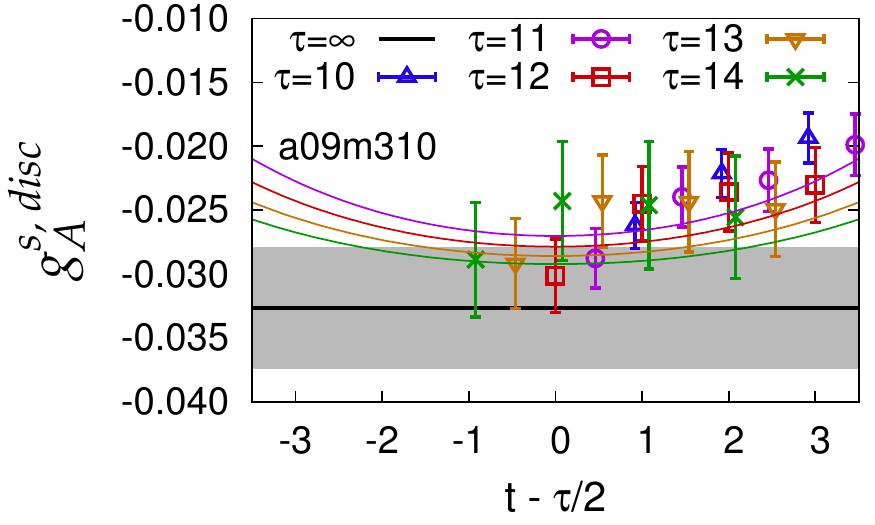}
    \includegraphics[height=1.0in,trim={1.08cm 0.70cm 0 0.06cm},clip]{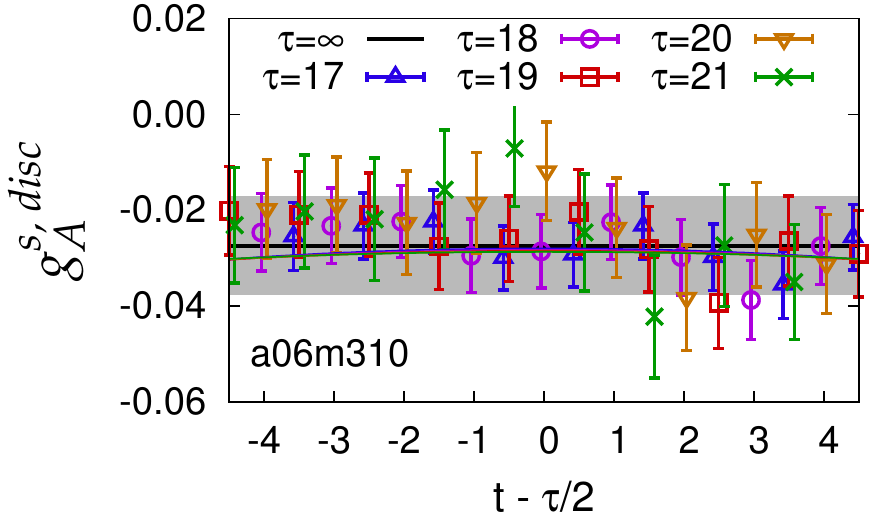}
  }\\
  \subfigure{
    \includegraphics[height=1.2in,trim={0.08cm 0.10cm 0 0.06cm},clip]{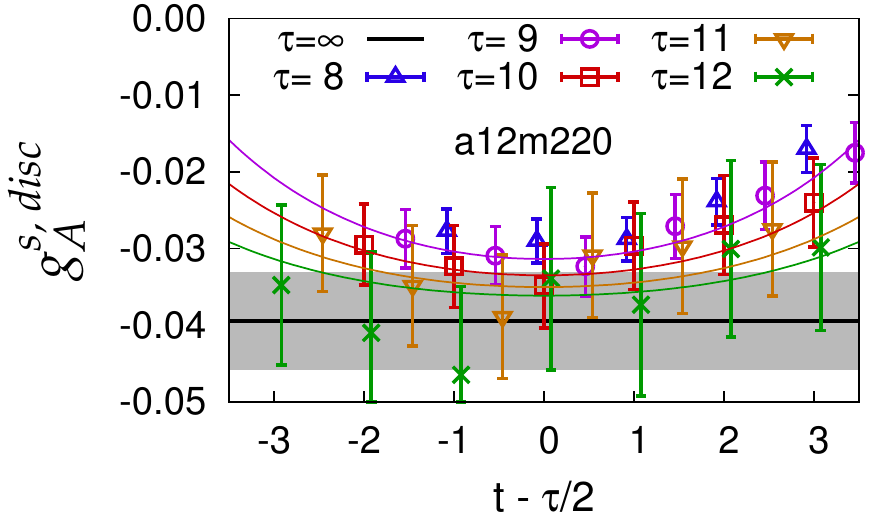}
    \includegraphics[height=1.2in,trim={1.08cm 0.10cm 0 0.06cm},clip]{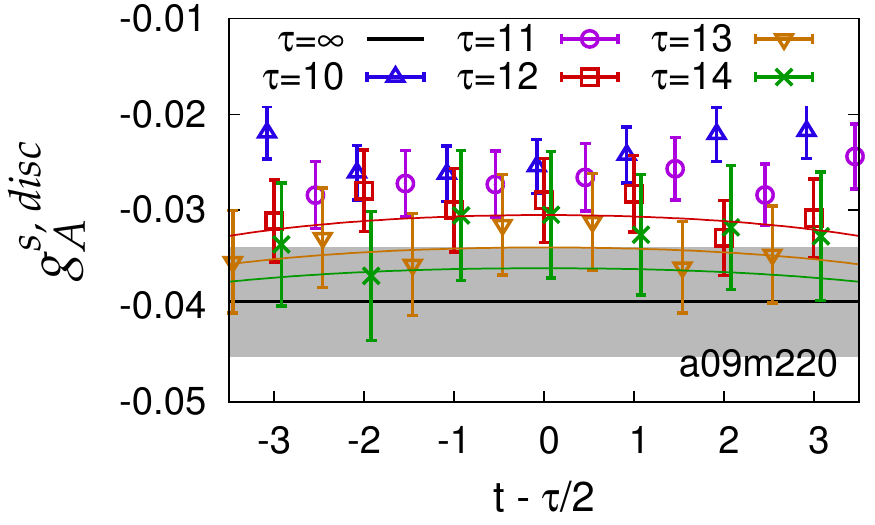}
    \includegraphics[height=1.2in,trim={1.08cm 0.10cm 0 0.06cm},clip]{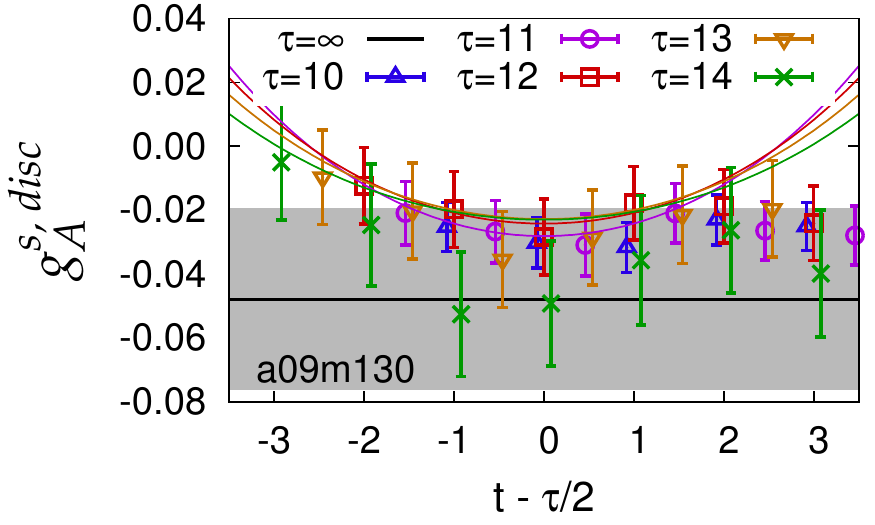}
  }
  \end{center}  
\vspace{-0.4cm}                                                                            
\caption{The data and the two-state fit to the light (top two rows) and strange (bottom two rows) quark disconnected
  contribution to the bare $g_A^{(l,s), {\text disc}}$. The grey error band and the solid line within it
  is the $\tau \to \infty$ estimate obtained using the two-state fit to
  data at different $t$ and $\tau$. The result of the fit for each
  individual $\tau$ is shown by a solid line in the same color as the
  data points.
  \label{fig:gAdisc}}
\end{figure*}

\section{Excited-State Contamination}
\label{sec:ESC}

To obtain the nucleon charges, we need to evaluate the matrix elements
of the corresponding quark bilinear operators within the ground state
of the nucleons.  We use the same toolkit to remove the excited-state
contamination (ESC) that is described in
Refs.~\cite{Bhattacharya:2015wna,Gupta:2018qil}: three-state
(two-state) fits to data for the connected (disconnected) three-point
functions as a function of both the operator insertion time $t$ and
multiple source-sink separation $\tau$.  The overlap with the ground
state is increased by using Gaussian smeared sources for propagator
calculation.  The root-mean-square smearing radius is tuned to be
between 0.6 and 0.7~fm. All correlation functions are constructed using
these propagators smeared at the source and the sink. The masses and
amplitudes of the states are extracted from the two-point functions
constructed using four-state fits. The details of these analyses have
already been published in Ref.~\cite{Gupta:2018qil}.  \looseness-1

The data and the two-state fits for the disconnected contributions are
shown in Fig.~\ref{fig:gAdisc}. The data are noisier compared to the
connected part analyzed in Ref.~\cite{Gupta:2018qil}.  Because of the
weaker statistical signal, the two-state fits to the three-point
function are, in some cases, more weighted towards smaller values of
the source-sink separation $\tau$. Also, in many cases there is no
clear pattern of convergence towards the $\tsepi$ value. We,
therefore, first determined the direction of convergence versus $\tau$
for both $g_{A}^{l}$ and $g_{A}^{s}$ by analyzing data at small $\tau$
that have smaller statistical errors but larger ESC. We then take the
largest range of $\tau$, shown in Fig.~\ref{fig:gAdisc}, for which the
errors are reasonable and the entries in the covariance matrix used in
the two-state fits are stable under variation in the set of values of
$t$ and $\tau$ used. Because of the difference in the quality of the
statistical signal, and because the number of ensembles and
configurations analyzed are not the same, we carry out separate
analyses of the connected and disconnected contributions.

Analyzing the connected and disconnected contributions separately to
remove ESC introduces an approximation. To define connected and
disconnected contributions individually, one has to work in a
partially quenched theory with an additional quark with flavor
$u^\prime$. However, in this theory the Pauli exclusion principle does
not apply between the $u$ and $u^\prime$ quarks.  The upshot of this
is that the spectrum of states in the partially quenched theory is
larger, for example, an intermediate $u^\prime u d$ state would be the
analogue of a $\Lambda$ rather than a
nucleon~\cite{Sharpe:2018PQ}. Thus, the spectral decomposition for
this partially quenched theory and QCD is different. In the ESC fits,
we however use the same QCD spectral decomposition in the fits for
both the two- and three-point functions, whereas one should be using
the partially quenched spectrum for the three-point function.  The
size of the extrapolation under consideration is the difference
between the value at $t=\tau/2$ for the largest $\tau$ and the
asymptotic value, whose estimate is the grey band. Since this
difference should converge exponentially as $\tau \to \infty$ and is
observed to be small ($< 0.02$), as shown in Fig.~\ref{fig:gAdisc}, we
assume that any additional systematic in the extrapolation due to not
using the partially quenched spectrum is well within the quoted
uncertainty.  We have also found that in the fits to both connected
and disconnected contributions, the extrapolated value is not very
sensitive to the precise values of the amplitudes and masses used, i.e., whether 
they are taken from three- or four-state fits to the two-point functions.  
\looseness-1

In the fits to remove ESC, we underscore the observation that the
disconnected contribution converges from above, while the connected
contribution converges from below as shown in
Ref.~\cite{Gupta:2018qil}; i.e., the ESC in the disconnected
three-point function is opposite to that observed in the connected
three-point function. In both cases, removing the ESC increases
the magnitude of their contribution, with the disconnected contribution 
becoming more negative. Note that such an increase in the negative 
contribution from the sea quarks reduces the fraction of the
nucleon spin carried by the quarks.\looseness-1

The final results for the bare values of the light ($u,d$), $g_A^{l,
  {\rm bare}}$, and strange quarks, $g_A^{s, {\rm bare}}$, obtained
from the two-state fits are collected together in
Table~\ref{tab:results}. \looseness-1

\begin{figure*}[th]
\centering
  \subfigure{
    \includegraphics[height=1.18in,trim={0.1cm   0.10cm 0 0.1cm},clip]{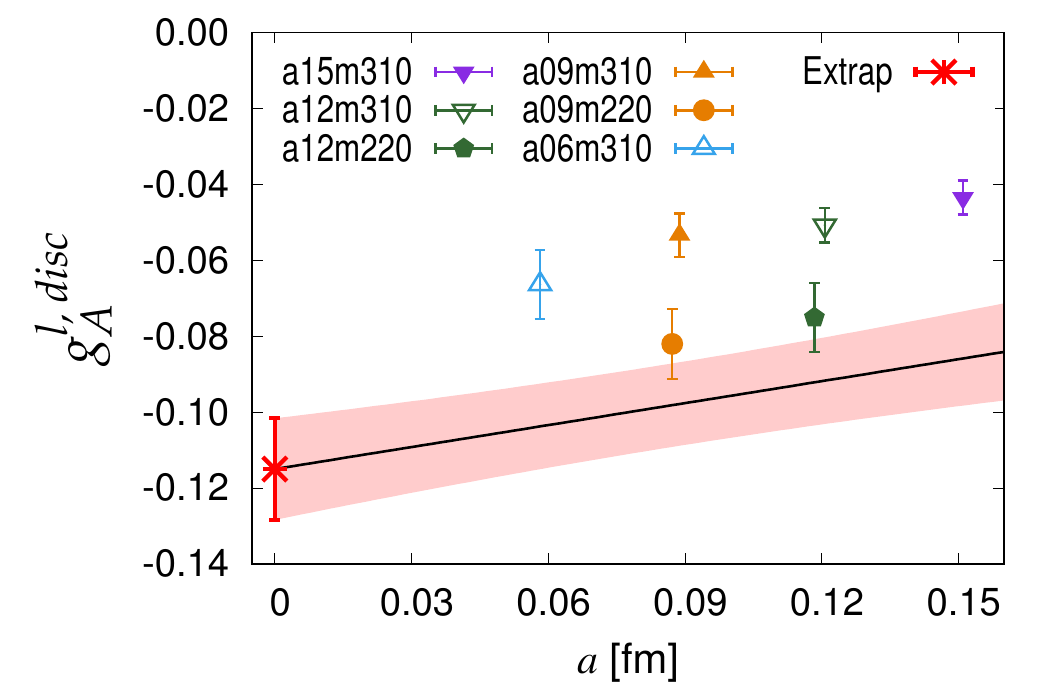} 
    \includegraphics[height=1.18in,trim={1.3cm   0.10cm 0 0.1cm},clip]{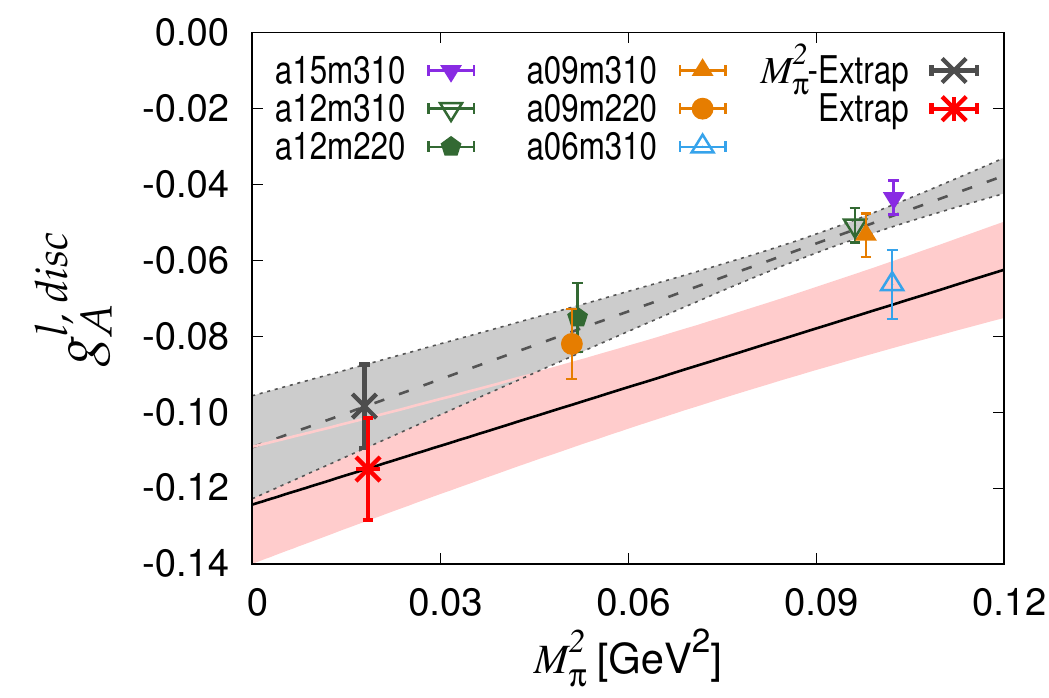}
    \includegraphics[height=1.18in,trim={0.1cm   0.10cm 0 0.1cm},clip]{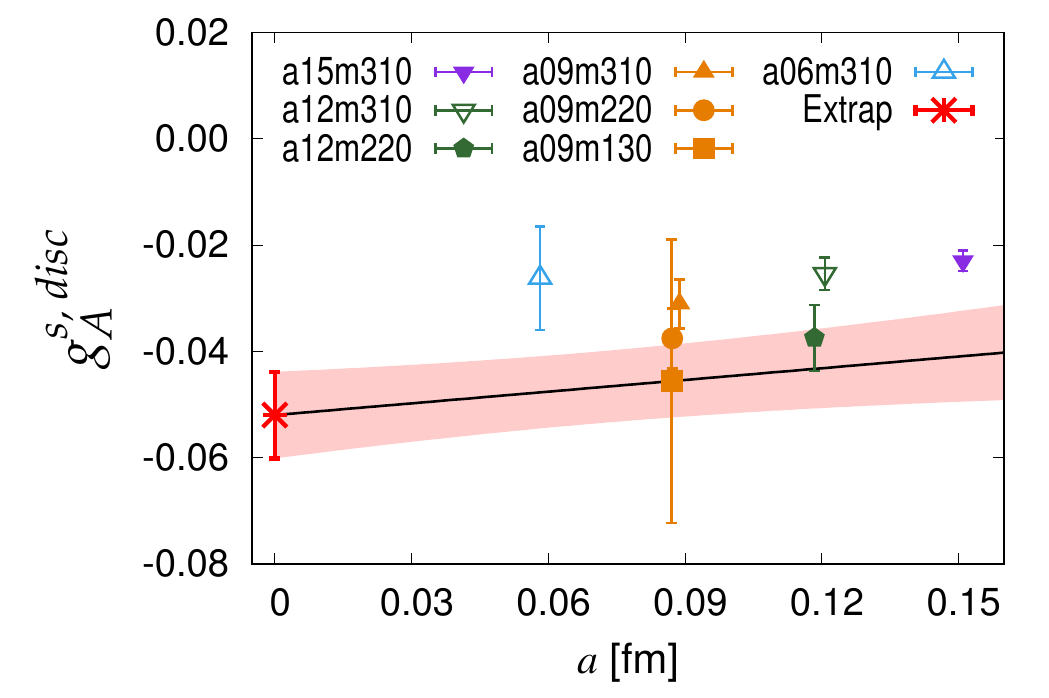} 
    \includegraphics[height=1.18in,trim={1.3cm   0.10cm 0 0.1cm},clip]{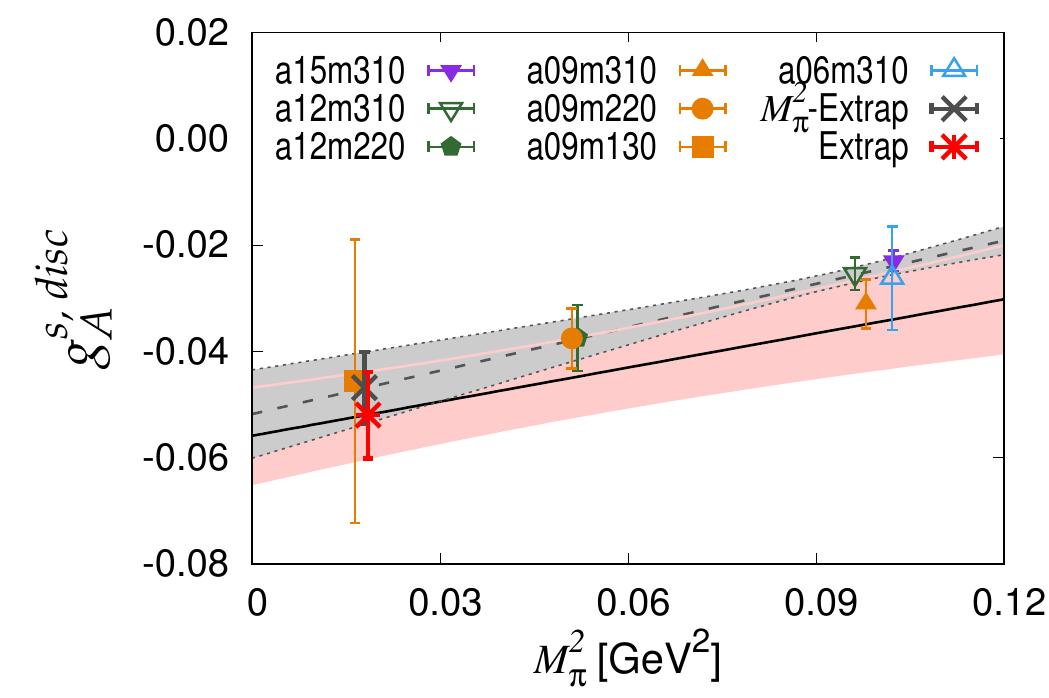} 
  }
  \subfigure{
    \includegraphics[height=1.18in,trim={0.1cm   0.10cm 0 0.1cm},clip]{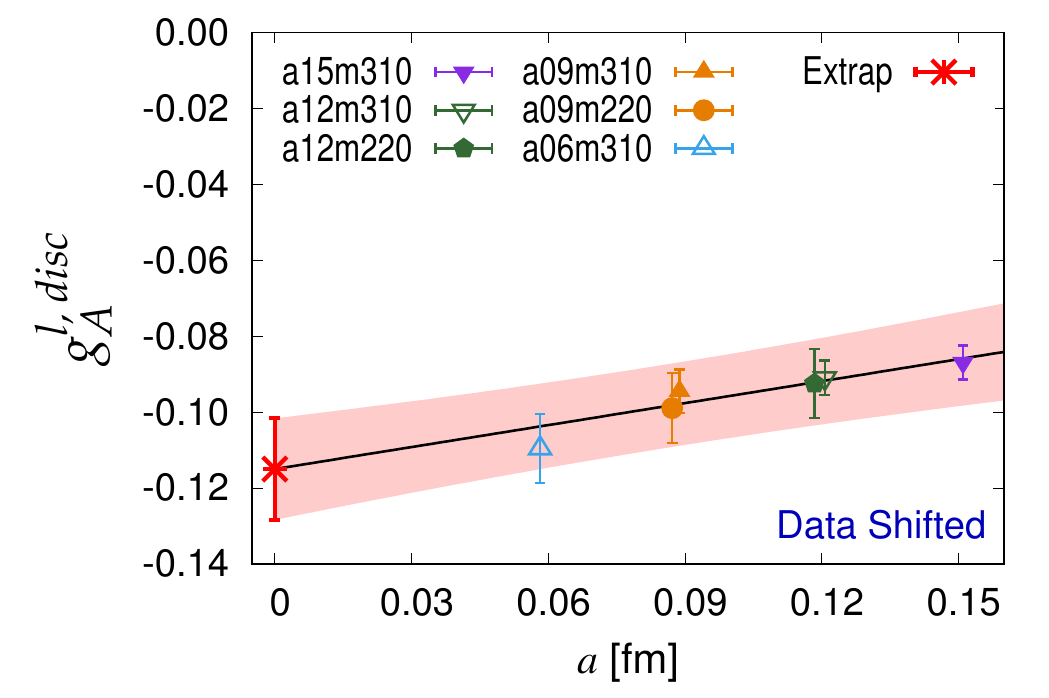} 
    \includegraphics[height=1.18in,trim={1.3cm   0.10cm 0 0.1cm},clip]{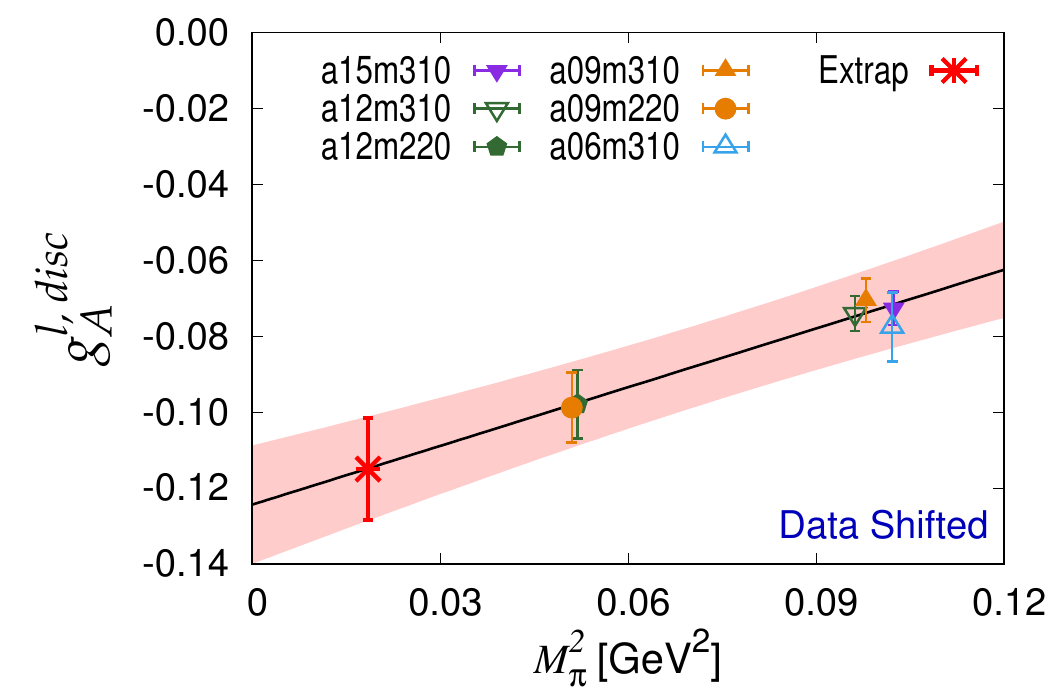}
    \includegraphics[height=1.18in,trim={0.1cm   0.10cm 0 0.1cm},clip]{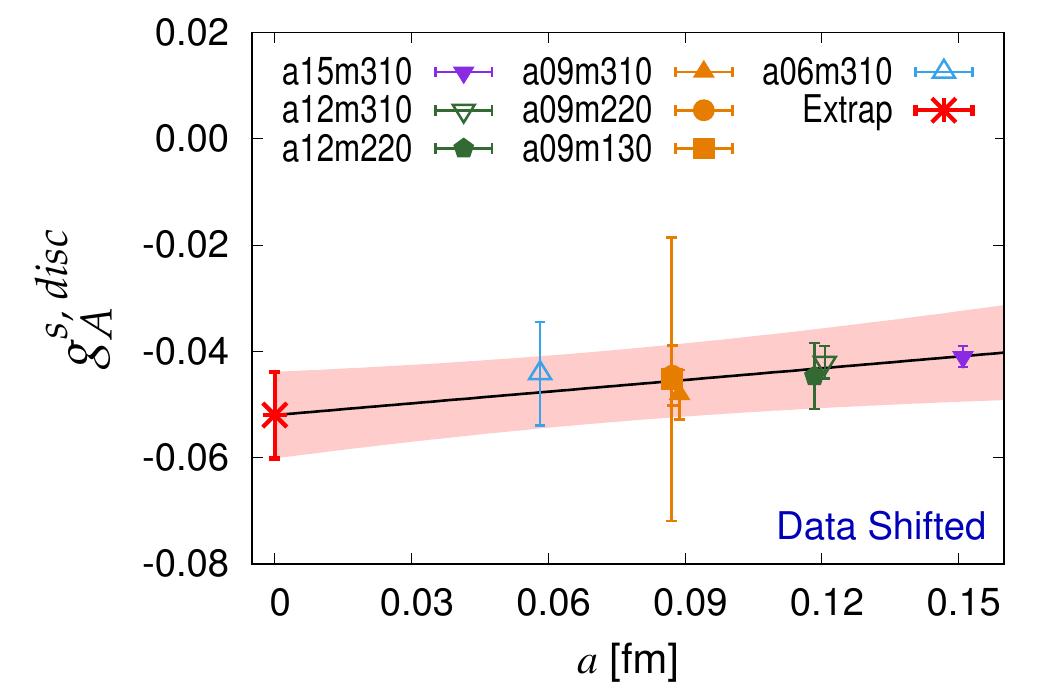} 
    \includegraphics[height=1.18in,trim={1.3cm   0.10cm 0 0.1cm},clip]{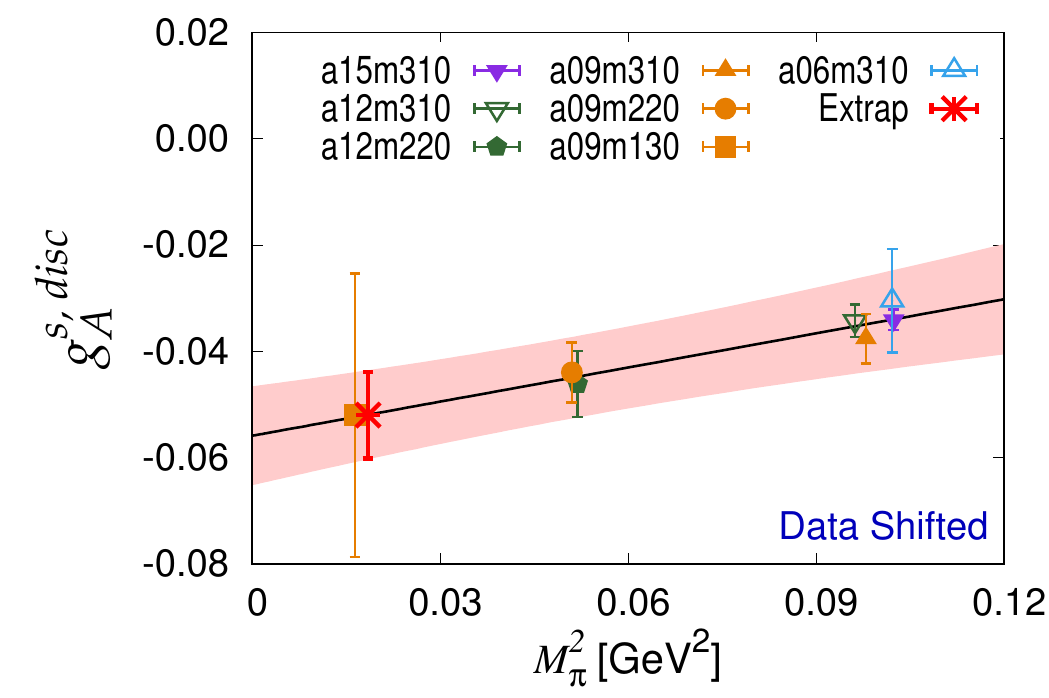} 
  }
\vspace{-0.3cm}
\caption{(Top) The extrapolation of the renormalized $g_A^{l,{\rm
      disc}}$ and $g_A^{s,{\rm disc}}$ data using the chiral-continuum
  ansatz given in Eq.~\protect\eqref{eq:extrapgAST}.  In each panel,
  the pink band shows the result of the simultaneous fit plotted
  versus a single variable with the other variable set to its physical
  value. The result at the physical point, $M_\pi = 135$~MeV and
  $a=0$, is marked with a red star. The grey band shows the fit versus
  only $M_\pi$, i.e., ignoring the dependence on $a$. It highlights
  the need for a simultaneous fit in both $a$ and $M_\pi$.  (Bottom)
  The data in each panel are plotted after extrapolation to the
  physical point ($a=0$ or $M_\pi=135$~MeV) in the nonplotted
  variable to facilitate comparison with the simultaneous
  fit. \looseness-1
  \label{fig:gls-extrap}}
\end{figure*}
%

\section{Renormalization of the operators}
\label{sec:renorm}

The renormalization of flavor diagonal light quark operators,
$\overline{q} \gamma_\mu \gamma_5 q$, requires knowing both nonsinglet
and singlet factors~\cite{Bhattacharya:2005rb}. In this work, we
neglect the difference between the two and renormalize all charges
using $Z_A^{\rm isovector}$ calculated in the
regularization-independent symmetric momentum-subtraction (RI-sMOM)
scheme and converted to the MS scheme at 2 GeV using two-loop
perturbation theory. These results are given in
Ref.~\cite{Gupta:2018qil}.  In perturbation theory, the difference
between the two starts at two loops as shown in
Ref.~\cite{Constantinou:2016ieh}; however the numerical value is
small, $O(0.01)$. Explicit nonperturbative calculations find that
$Z_A^{\rm nonsinglet}$ and $Z_A^{\rm singlet}$ agree to within a
percent for the twisted mass and the clover-Wilson
actions~\cite{Alexandrou:2017qyt,Alexandrou:2017oeh,Green:2017keo}.
While we have not checked that the difference is similarly small also
for our clover-on-HISQ calculation, we assume it is covered by the
$O(0.03)$ uncertainty in the calculated values of $Z_A^{\rm
  isovector}\equiv Z_A^{u-d}$ used.  In short, both the disconnected
and connected contributions are renormalized in the following two
ways:
\begin{align}
g_A|_{R1} &= g_A \times Z_A^{\rm isovector} \,, \nonumber \\
g_A|_{R2} &= \frac{g_A}{g_V^{u-d}} \times  \frac{Z_A^{\rm isovector}}{ Z_V^{u-d}}   \,. 
\label{eq:renorm}
\end{align}
with the values of $Z_A^{\rm isovector}$ and $Z_V^{u-d}$ taken from Ref.~\cite{Gupta:2018qil}. 
In the second method, the conserved vector current relation $g_V^{u-d} \times
Z_V^{u-d} = 1$ is implicit.  The final values are taken to be the average of the
two after performing the chiral-continuum extrapolation. The results
for disconnected contributions are given in Table~\ref{tab:results}
and the connected contributions, taken from Ref.~\cite{Gupta:2018qil},
are reproduced in Eq.~\eqref{eq:gAconn}. \looseness-1

\section{The Continuum-Chiral Extrapolation}
\label{sec:CCFV}

The leading discretization effects are taken to be linear in $a$ since
the action and the operators in our clover-on-HISQ formalism are not
fully $O(a)$ improved.  We take the leading dependence on $M_\pi$ from
the finite volume chiral perturbation
theory~\cite{Bernard:1992qa,Bernard:1995dp,Bernard:2006gx,Bernard:2006te,Khan:2006de,Colangelo:2010ba,deVries:2010ah} 
that is proportional to $M_\pi^2$.  We neglect finite volume
corrections since no significant evidence for them was found in the
dominant connected contributions~\cite{Gupta:2018qil}. In
Fig.~\ref{fig:gls-extrap}, we show the simultaneous chiral-continuum
fits (pink band) versus $a$ and $M_\pi^2$ to the renormalized disconnected data
$g_A^{l,s}$ given in Table~\ref{tab:results} using the
ansatz,\looseness-1
\begin{align}
  g_{A}^{l,s} (a,M_\pi,L) = c_1 + c_2 a + c_3 M_\pi^2  \,.
\label{eq:extrapgAST} 
\end{align}
The results of the extrapolated values from the fits for both
renormalization procedures are also given in Table~\ref{tab:results}
along with the $\chi^2/$DOF. For comparison, the grey band within
dotted lines in Fig.~\ref{fig:gls-extrap} is the fit to a single
variable $M_\pi^2$, i.e., with $c_2=0$.  The difference between the two bands 
highlights the need for the simultaneous fit. 

The results for the fit parameters $c_i$ for the light and strange
quarks are given in Table~\ref{tab:parameters}. From the fits shown in
Fig.~\ref{fig:gls-extrap}, it is clear that, even with the limited
number of data points, the signal good enough to give a statistically
significant determination of the $c_i$, and show that the variation is
essentially linear in the two variables. The change in the $c_i$ on
going from the light to the strange quark is also clear from the data and the fits.

\begin{table}[h]   
\centering
\begin{ruledtabular}
\begin{tabular}{c|cc|cc}
                   &  $g_A^{l}|_{R1}$  & $g_A^{s}|_{R1}$   & $g_A^{l}|_{R2}$ & $g_A^{s}|_{R2}$  \\ 
\hline                                                                                                     
$c_1$              & $-$0.124(15)      & $-$0.056(9)       & $-$0.129(15)    &  $-$0.058(9)   \\
$c_2$ (fm$^{-1}$)  &    0.193(89)      &    0.073(65)      &    0.207(88)    &     0.084(65)  \\
$c_3$ (GeV$^{-2}$) &    0.52(15)       &    0.21(10)       &    0.53(15)     &     0.22(10)   \\
$\chi^2/DOF$       &    0.281          &    0.167          &    0.203        &     0.205      \\
\end{tabular}
\end{ruledtabular}
\caption{The values of the parameters $c_i$, defined in
  Eq.~\eqref{eq:extrapgAST}, for the final fits shown in
  Fig.~\protect\ref{fig:gls-extrap} and the results given in
  Table~\protect\ref{tab:results}. \looseness-1}
\label{tab:parameters}
\end{table}

We also carried out fits including the next order corrections, $a^2$
for the discretization errors and $M_\pi^2 \log M_\pi^2$ for the
chiral log term, one at a time.  In each case, the errors in the
coefficients and in the results grow. For example, in the best case of
adding the $a^2$ term as there are data at four values of $a$, we get
$g_A^{l}|_{R2}= -0.147(43)$, and the coefficients $c_2=0.67(71)$ and
$c_{a^2}=-2.1(3.2)$. The $\chi^2=0.6$ of the fit using
Eq.~\eqref{eq:extrapgAST}, which was already unreasonably small,
decreased to 0.3. There was no scope to reduce $\chi^2$ by two units
as is required by the Akaike Information
Criteria~\protect\cite{1100705} to warrant including additional terms.
In fact, as is obvious, such statistical tests are meaningless for
such small $\chi^2$ values. More importantly, within the range of the
data, $c_{2} $ and $c_{a^2}$ compete to reduce $\chi^2$ but both are
poorly determined. Outside, the predictive power of the fit
deteriorates as is typical of overparametrized fits. Our conclusion is
that the ansatz given in Eq.~\eqref{eq:extrapgAST} is sufficient to
fit the current data and many more data points are needed to explore
additional corrections.  Unfortunately, the analysis of the remaining
HISQ ensembles at smaller $a$, and those at the physical pion mass,
has not yet been possible due to the computational cost.  Our final
results are, therefore, derived from fits using
Eq.~\eqref{eq:extrapgAST}. To account for the uncertainty in the fit
model, we assign an additional systematic error of $0.03$, coming from
the connected contributions, in $g_A^u$ and $g_A^d$.

Results for the individual contributions are collected together in
Table~\ref{tab:resultsFINAL}, along with the connected contributions
reproduced from Ref.~\cite{Gupta:2018qil}.  Their sum $\frac{1}{2}
\Delta \Sigma \equiv \sum_{q=u,d,s} \frac{1}{2} {\Delta q} =
0.143(31)$ is in good agreement with the COMPASS
result~\cite{Adolph:2015saz}. Scaling our value of $g_A^s$ by $1/m_q$
suggests that the neglected charm contribution could be 
$g_A^c \approx -0.005$. \looseness-1

Our result $g_A^{u-d} = 1.218(27)(30)$ for the isovector axial charge
is 0.058 below the experimental value $g_A^{u-d} = 1.2766(20)$, as
discussed in Ref.~\cite{Gupta:2018qil}. The difference can be
explained if the connected $g_A^u$ is underestimated by 0.058 or
$g_A^d$ is more negative by this amount or any combination of the
two. (The disconnected contributions cancel in $g_A^{u-d}$.) In the
first case, $ \Delta \Sigma/2$ would increase by 0.029, and in the
second case decrease by the same amount. The most likely reason for this
underestimate is the uncertainty in the chiral-continuum-finite-volume
extrapolation, which was estimated to be 0.03 in Ref. [8], independent
of the experimental value.  This systematic uncertainty has been added
as a second error in $g_A^{u}$ and $g_A^{d}$. In the fits to the
disconnected data shown in Fig 3, we do not find large deviations from
linearity. The quoted error is comparable to the change, $\approx 0.02$, between the
lowest $M_\pi$ and $a$ point and the extrapolated value (red star),
and thus a conservative estimate of possible
residual uncertainty. Another estimate of the same systematic, $a^2
\Lambda_{\rm QCD}^2 \approx 0.02$, gives a similar value. We do include $0.02$,
as an additional systematic for the sum of the disconnected
contributions; combine it in quadrature with that from the connected
contribution; and quote an overall second error of 0.036 in $ \Delta
\Sigma/2 = 0.143(31)(36)$. This represents the model uncertainty of
the chiral-continuum ansatz, i.e., using the lowest order corrections
and fitting to a limited number of data points.  \looseness-1

\begin{table}[t]
\centering
\begin{ruledtabular}
\begin{tabular}{c|ccc}
              & $g_A^{u} \equiv {\Delta u} $         & $g_A^{d} \equiv {\Delta d} $ & $g_A^{s} \equiv {\Delta s}$     \\ 
\hline
Connected     &    0.895(21)(30) & $-$0.320(12)(30)  &               \\
Disconnected  & $-$0.118(14)     & $-$0.118(14)      & $-$0.053(8)   \\ 
Sum           &    0.777(25)(30) & $-$0.438(18)(30)  & $-$0.053(8)   \\
\hline
ETMC          &    0.830(26)     & $-$0.386(18)      & $-$0.042(10)(2)  \\
\end{tabular}
\end{ruledtabular}
\caption{Our results for the $u,\ d$, and $s$ quarks, after extrapolation to $a=0$ and
  $M_\pi=135$~MeV, for the connected and disconnected contributions
  and their sum are given in the
  first three rows. The sum over flavors gives $\Delta \Sigma = \Delta u + \Delta d +
  \Delta s =0.286(62)$.  ETMC results at the single lattice spacing
  $a=0.0938$~fm~\protect\cite{Alexandrou:2017oeh} are given in the
  last row. \looseness-1}
\label{tab:resultsFINAL}
\end{table}

\section{Comparison with Previous Work and Conclusions}
\label{sec:comparison}

\begin{figure*}[tbh]
\begin{flushleft}                                               
  \subfigure{
    \includegraphics[height=1.5in,trim={0.15cm   0.15cm 0 0.1cm},clip]{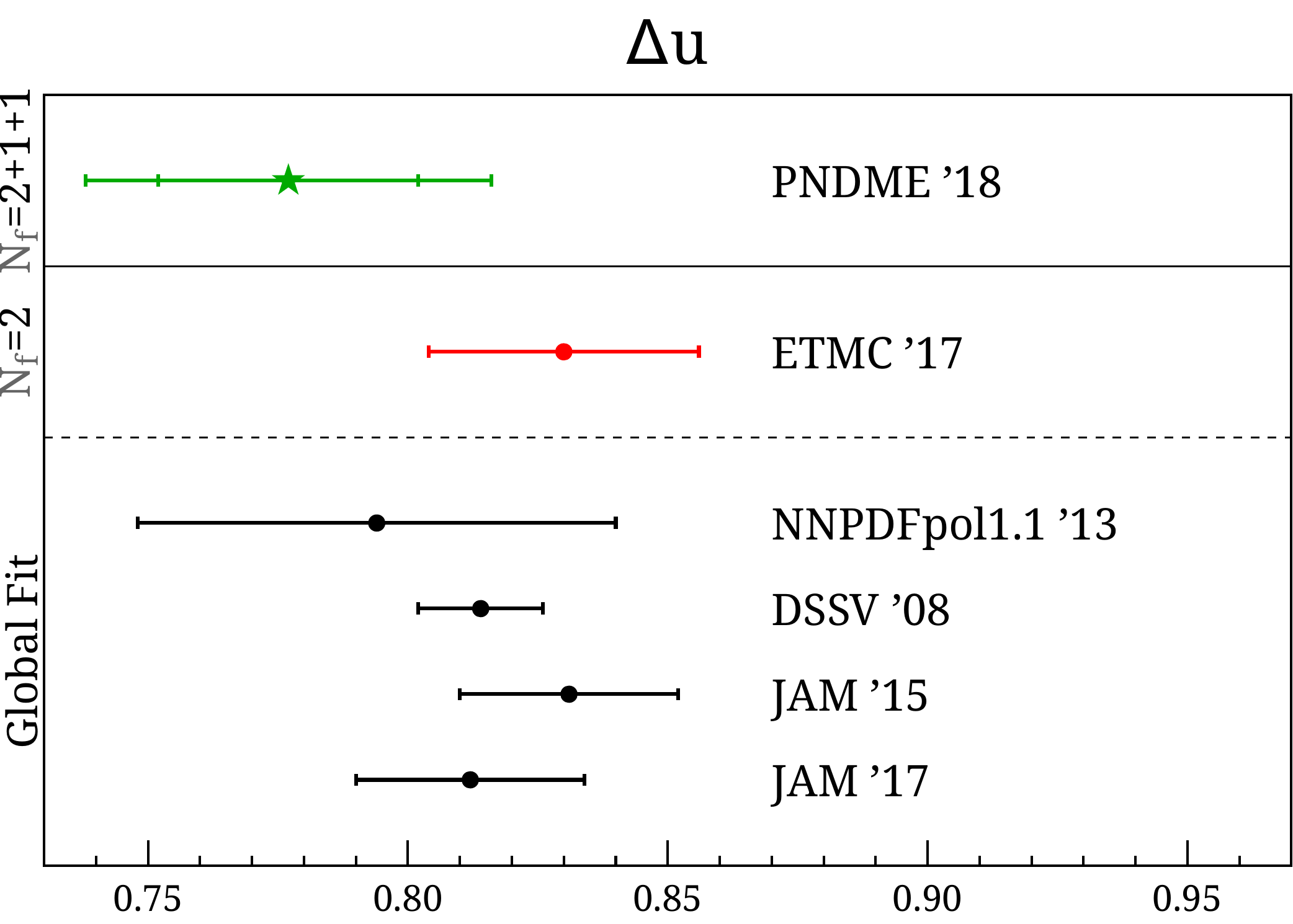} 
    \includegraphics[height=1.53in,trim={0.4cm   0.25cm 0 0.1cm},clip]{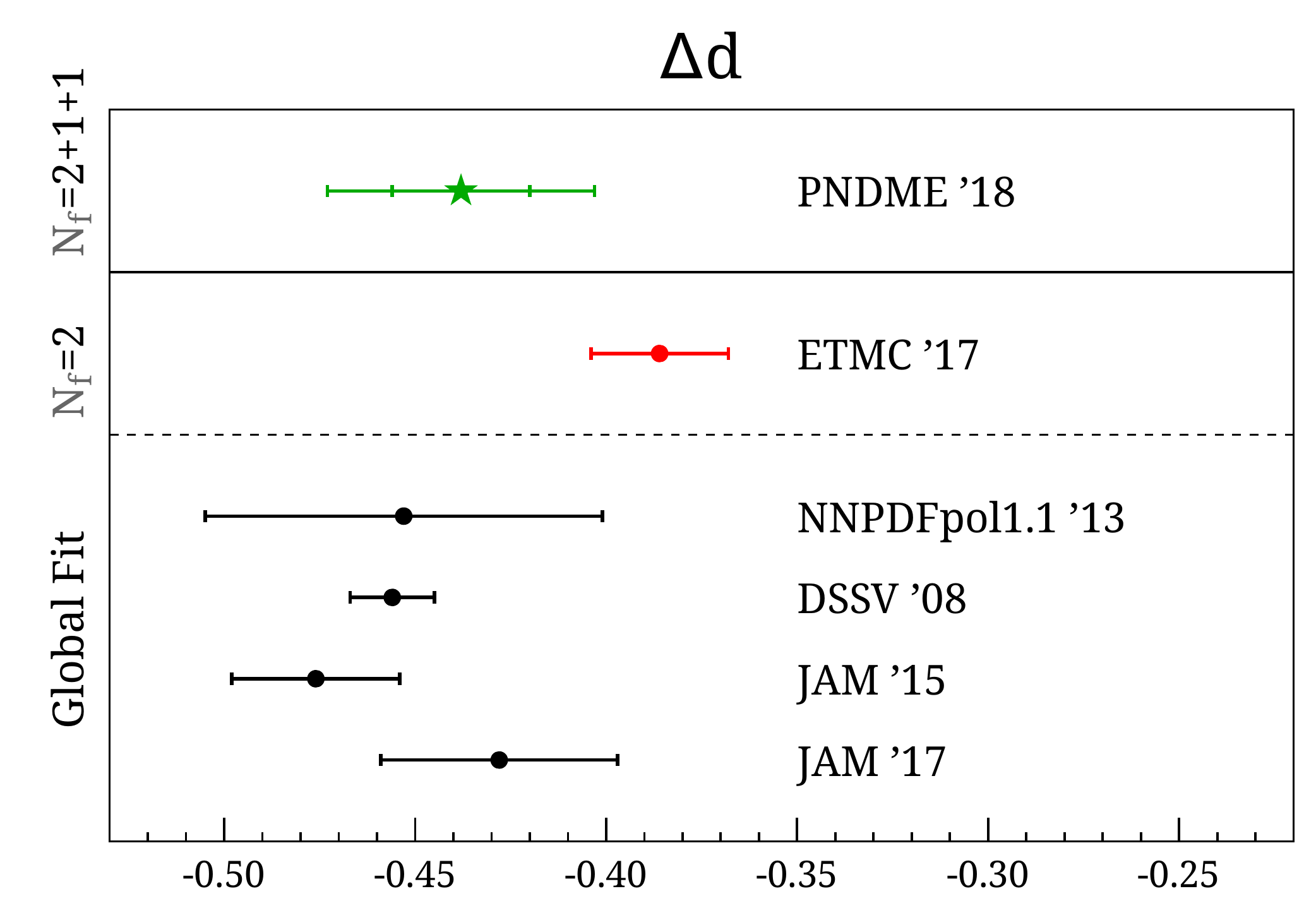} 
    \includegraphics[height=1.5in,trim={0.2cm    0.15cm 0 0.1cm},clip]{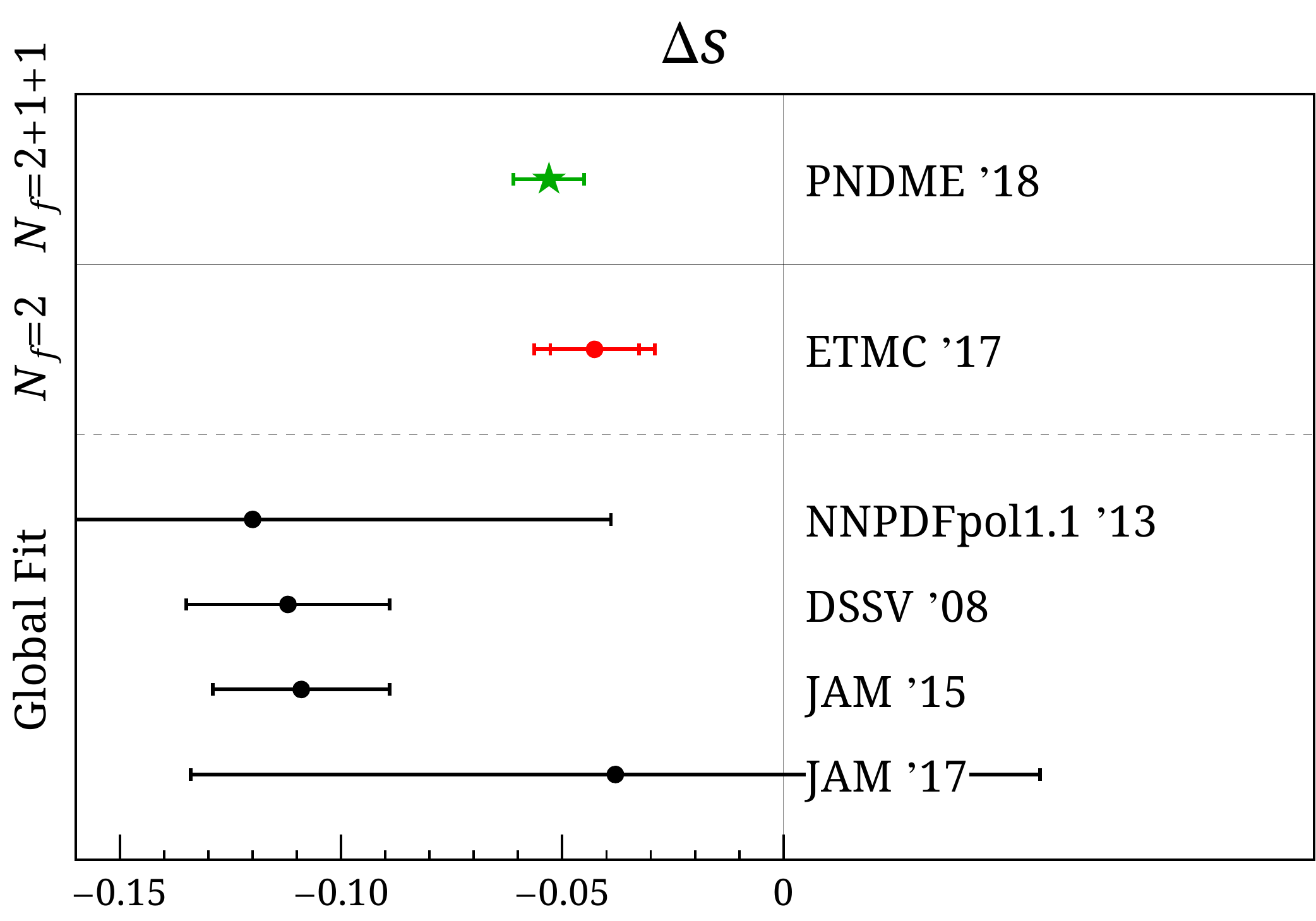} 
  }
 \end{flushleft}                                                                                            
\caption{Chiral-continuum extrapolated PNDME'18 results (this work)
  for $\Delta u$, $\Delta d$ and $\Delta s$ are compared with
  ETMC~\cite{Alexandrou:2017oeh,Alexandrou:2018lvq} values obtained at
  a single-lattice spacing and with moments from global fits to
  polarized PDF (NNPDFpol1.1'13~\protect\cite{Nocera:2014gqa},
  DSSV'08~\protect\cite{deFlorian:2008mr,deFlorian:2009vb},
  Jam'15~\protect\cite{Sato:2016tuz}, and
  JAM'17~\protect\cite{Ethier:2017zbq}). All PDF results are taken 
  from  Ref.~\protect\cite{Lin:2017snn} and are at 2~GeV in the $\overline{MS}$ scheme. 
\looseness-1
  \label{fig:PDF}}
\end{figure*}

In Fig.~\ref{fig:PDF}, we compare lattice results, restricted to
publications including physical mass ensembles, with the moments
extracted from global fits to the polarized PDFs reviewed in
Ref.~\protect\cite{Lin:2017snn}. Within errors, our results are
compatible with the moments extracted from global PDF fits, all
expressed in the $\overline{MS}$ scheme at 2~GeV.  The ETMC lattice
results from a single physical mass 2-flavor ensemble at
$a=0.093$~fm~\cite{Alexandrou:2017oeh,Alexandrou:2018lvq} are also
consistent with ours. The small difference can be accounted for by the
$a$ dependence highlighted in our disconnected contribution data shown
in Fig.~\ref{fig:gls-extrap}, i.e., to get continuum limit values
assuming similar discretization errors, our fits indicate subtracting
$0.04$ from their $g_A^u$ and $g_A^d$ and $0.01$ from $g_A^s$. The
change in the connected contributions to $g_A^u$ and $g_A^d$
is only $O(0.01)$~\cite{Gupta:2018qil}. Most likely, this is because the ETMC
value for the isovector charge $g_A^{u-d}=1.212(40)$ is equally low.
The difference in the two respective error estimates is mainly due to
ETMC not including a systematic uncertainty to account for possible
discretization effects in the connected and disconnected
contributions, and therefore in $\Delta \Sigma$.  \looseness-1

In conclusion, we present first results with chiral-continuum
extrapolation of up, down and strange quark spin contributions.  These
fits are based on 6 (7) ensembles for the disconnected contribution of
light (strange) quarks, and on 11 ensembles for the dominant connected
contributions that were analyzed fully in Ref.~\cite{Gupta:2018qil}.
We demonstrate in Fig.~\ref{fig:gls-extrap} that a chiral-continuum
extrapolation significantly reduces the disconnected contribution of
the quark spin to the proton spin and is, therefore, essential for
getting physical results from lattice calculations. Our final result,
$\frac{1}{2} {\Delta \Sigma} = 0.143(31)(36)$, is consistent with the
2015 COMPASS analysis. Including more ensembles in the calculation of
the disconnected contribution and resolving the $\approx 5\%$
underestimate of $g_A^{u-d}$ that impacts the connected contributions
are, at present, the largest systematics that need to be addressed in
future works.\looseness-1

\vspace{0.4cm}
{\it Acknowledgments}: We thank the MILC Collaboration for providing the 2+1+1-flavor HISQ
lattices. The calculations used the Chroma
software suite~\cite{Edwards:2004sx}. Simulations were carried out on
computer facilities at (i) the National Energy Research Scientific
Computing Center, a DOE Office of Science User Facility supported by
the Office of Science of the U.S. Department of Energy under Contract
No. DE-AC02-05CH11231; and, (ii) the Oak Ridge Leadership Computing
Facility at the Oak Ridge National Laboratory, which is supported by
the Office of Science of the U.S. Department of Energy under Contract
No. DE-AC05-00OR22725; (iii) the USQCD Collaboration, which are funded
by the Office of Science of the U.S. Department of Energy, (iv)
Institutional Computing at Los Alamos National Laboratory,
and (v) Institute for Cyber-Enabled Research at Michigan State University. 
T. Bhattacharya and R. Gupta were partly supported by the
U.S. Department of Energy, Office of Science, Office of High Energy
Physics under Contract No.~DE-AC52-06NA25396.  T. Bhattacharya,
R. Gupta, Y-C. Jang and B. Yoon were partly supported by
the LANL LDRD program.  The work of H.-W. Lin is supported by the US National
Science Foundation under grant PHY 1653405 ``CAREER: Constraining
Parton Distribution Functions for New-Physics Searches''. \looseness-1

\clearpage
%
\bibliography{ref} 

\begin{thebibliography}{33}%
\makeatletter
\providecommand \@ifxundefined [1]{%
 \@ifx{#1\undefined}
}%
\providecommand \@ifnum [1]{%
 \ifnum #1\expandafter \@firstoftwo
 \else \expandafter \@secondoftwo
 \fi
}%
\providecommand \@ifx [1]{%
 \ifx #1\expandafter \@firstoftwo
 \else \expandafter \@secondoftwo
 \fi
}%
\providecommand \natexlab [1]{#1}%
\providecommand \enquote  [1]{``#1''}%
\providecommand \bibnamefont  [1]{#1}%
\providecommand \bibfnamefont [1]{#1}%
\providecommand \citenamefont [1]{#1}%
\providecommand \href@noop [0]{\@secondoftwo}%
\providecommand \href [0]{\begingroup \@sanitize@url \@href}%
\providecommand \@href[1]{\@@startlink{#1}\@@href}%
\providecommand \@@href[1]{\endgroup#1\@@endlink}%
\providecommand \@sanitize@url [0]{\catcode `\\12\catcode `\$12\catcode
  `\&12\catcode `\#12\catcode `\^12\catcode `\_12\catcode `\%12\relax}%
\providecommand \@@startlink[1]{}%
\providecommand \@@endlink[0]{}%
\providecommand \url  [0]{\begingroup\@sanitize@url \@url }%
\providecommand \@url [1]{\endgroup\@href {#1}{\urlprefix }}%
\providecommand \urlprefix  [0]{URL }%
\providecommand \Eprint [0]{\href }%
\providecommand \doibase [0]{http://dx.doi.org/}%
\providecommand \selectlanguage [0]{\@gobble}%
\providecommand \bibinfo  [0]{\@secondoftwo}%
\providecommand \bibfield  [0]{\@secondoftwo}%
\providecommand \translation [1]{[#1]}%
\providecommand \BibitemOpen [0]{}%
\providecommand \bibitemStop [0]{}%
\providecommand \bibitemNoStop [0]{.\EOS\space}%
\providecommand \EOS [0]{\spacefactor3000\relax}%
\providecommand \BibitemShut  [1]{\csname bibitem#1\endcsname}%
\let\auto@bib@innerbib\@empty
\bibitem [{\citenamefont {Ashman}\ \emph {et~al.}(1988)\citenamefont {Ashman}
  \emph {et~al.}}]{Ashman:1987hv}%
  \BibitemOpen
  \bibfield  {author} {\bibinfo {author} {\bibfnamefont {J.}~\bibnamefont
  {Ashman}} \emph {et~al.} (\bibinfo {collaboration} {European Muon}),\
  }\bibfield  {booktitle} {\emph {\bibinfo {booktitle} {{Internal spin
  structure of the nucleon. Proceedings, Symposium, SMC Meeting, New Haven,
  USA, January 5-6, 1994}}},\ }\href {\doibase 10.1016/0370-2693(88)91523-7}
  {\bibfield  {journal} {\bibinfo  {journal} {Phys. Lett.}\ }\textbf {\bibinfo
  {volume} {B206}},\ \bibinfo {pages} {364} (\bibinfo {year}
  {1988})}\BibitemShut {NoStop}%
\bibitem [{\citenamefont {Adolph}\ \emph {et~al.}(2016)\citenamefont {Adolph}
  \emph {et~al.}}]{Adolph:2015saz}%
  \BibitemOpen
  \bibfield  {author} {\bibinfo {author} {\bibfnamefont {C.}~\bibnamefont
  {Adolph}} \emph {et~al.} (\bibinfo {collaboration} {COMPASS}),\ }\href
  {\doibase 10.1016/j.physletb.2015.11.064} {\bibfield  {journal} {\bibinfo
  {journal} {Phys. Lett.}\ }\textbf {\bibinfo {volume} {B753}},\ \bibinfo
  {pages} {18} (\bibinfo {year} {2016})},\ \Eprint
  {http://arxiv.org/abs/1503.08935} {arXiv:1503.08935 [hep-ex]} \BibitemShut
  {NoStop}%
\bibitem [{\citenamefont {Ji}(1997)}]{Ji:1996ek}%
  \BibitemOpen
  \bibfield  {author} {\bibinfo {author} {\bibfnamefont {X.-D.}\ \bibnamefont
  {Ji}},\ }\href {\doibase 10.1103/PhysRevLett.78.610} {\bibfield  {journal}
  {\bibinfo  {journal} {Phys. Rev. Lett.}\ }\textbf {\bibinfo {volume} {78}},\
  \bibinfo {pages} {610} (\bibinfo {year} {1997})},\ \Eprint
  {http://arxiv.org/abs/hep-ph/9603249} {arXiv:hep-ph/9603249 [hep-ph]}
  \BibitemShut {NoStop}%
\bibitem [{\citenamefont {Lin}\ \emph {et~al.}(2018)\citenamefont {Lin} \emph
  {et~al.}}]{Lin:2017snn}%
  \BibitemOpen
  \bibfield  {author} {\bibinfo {author} {\bibfnamefont {H.-W.}\ \bibnamefont
  {Lin}} \emph {et~al.},\ }\href {\doibase 10.1016/j.ppnp.2018.01.007}
  {\bibfield  {journal} {\bibinfo  {journal} {Prog. Part. Nucl. Phys.}\
  }\textbf {\bibinfo {volume} {100}},\ \bibinfo {pages} {107} (\bibinfo {year}
  {2018})},\ \Eprint {http://arxiv.org/abs/1711.07916} {arXiv:1711.07916
  [hep-ph]} \BibitemShut {NoStop}%
\bibitem [{\citenamefont {Fitzpatrick}\ \emph {et~al.}(2012)\citenamefont
  {Fitzpatrick}, \citenamefont {Haxton}, \citenamefont {Katz}, \citenamefont
  {Lubbers},\ and\ \citenamefont {Xu}}]{Fitzpatrick:2012ib}%
  \BibitemOpen
  \bibfield  {author} {\bibinfo {author} {\bibfnamefont {A.~L.}\ \bibnamefont
  {Fitzpatrick}}, \bibinfo {author} {\bibfnamefont {W.}~\bibnamefont {Haxton}},
  \bibinfo {author} {\bibfnamefont {E.}~\bibnamefont {Katz}}, \bibinfo {author}
  {\bibfnamefont {N.}~\bibnamefont {Lubbers}}, \ and\ \bibinfo {author}
  {\bibfnamefont {Y.}~\bibnamefont {Xu}},\ }\href@noop {} {\  (\bibinfo {year}
  {2012})},\ \Eprint {http://arxiv.org/abs/1211.2818} {arXiv:1211.2818
  [hep-ph]} \BibitemShut {NoStop}%
\bibitem [{\citenamefont {Hill}\ and\ \citenamefont
  {Solon}(2015)}]{Hill:2014yxa}%
  \BibitemOpen
  \bibfield  {author} {\bibinfo {author} {\bibfnamefont {R.~J.}\ \bibnamefont
  {Hill}}\ and\ \bibinfo {author} {\bibfnamefont {M.~P.}\ \bibnamefont
  {Solon}},\ }\href {\doibase 10.1103/PhysRevD.91.043505} {\bibfield  {journal}
  {\bibinfo  {journal} {Phys. Rev.}\ }\textbf {\bibinfo {volume} {D91}},\
  \bibinfo {pages} {043505} (\bibinfo {year} {2015})},\ \Eprint
  {http://arxiv.org/abs/1409.8290} {arXiv:1409.8290 [hep-ph]} \BibitemShut
  {NoStop}%
\bibitem [{\citenamefont {Bhattacharya}\ \emph {et~al.}(2015)\citenamefont
  {Bhattacharya}, \citenamefont {Cirigliano}, \citenamefont {Cohen},
  \citenamefont {Gupta}, \citenamefont {Joseph}, \citenamefont {Lin},\ and\
  \citenamefont {Yoon}}]{Bhattacharya:2015wna}%
  \BibitemOpen
  \bibfield  {author} {\bibinfo {author} {\bibfnamefont {T.}~\bibnamefont
  {Bhattacharya}}, \bibinfo {author} {\bibfnamefont {V.}~\bibnamefont
  {Cirigliano}}, \bibinfo {author} {\bibfnamefont {S.}~\bibnamefont {Cohen}},
  \bibinfo {author} {\bibfnamefont {R.}~\bibnamefont {Gupta}}, \bibinfo
  {author} {\bibfnamefont {A.}~\bibnamefont {Joseph}}, \bibinfo {author}
  {\bibfnamefont {H.-W.}\ \bibnamefont {Lin}}, \ and\ \bibinfo {author}
  {\bibfnamefont {B.}~\bibnamefont {Yoon}} (\bibinfo {collaboration} {PNDME}),\
  }\href {\doibase 10.1103/PhysRevD.92.094511} {\bibfield  {journal} {\bibinfo
  {journal} {Phys. Rev.}\ }\textbf {\bibinfo {volume} {D92}},\ \bibinfo {pages}
  {094511} (\bibinfo {year} {2015})},\ \Eprint
  {http://arxiv.org/abs/1506.06411} {arXiv:1506.06411 [hep-lat]} \BibitemShut
  {NoStop}%
\bibitem [{\citenamefont {Gupta}\ \emph {et~al.}(2018)\citenamefont {Gupta},
  \citenamefont {Jang}, \citenamefont {Yoon}, \citenamefont {Lin},
  \citenamefont {Cirigliano},\ and\ \citenamefont
  {Bhattacharya}}]{Gupta:2018qil}%
  \BibitemOpen
  \bibfield  {author} {\bibinfo {author} {\bibfnamefont {R.}~\bibnamefont
  {Gupta}}, \bibinfo {author} {\bibfnamefont {Y.-C.}\ \bibnamefont {Jang}},
  \bibinfo {author} {\bibfnamefont {B.}~\bibnamefont {Yoon}}, \bibinfo {author}
  {\bibfnamefont {H.-W.}\ \bibnamefont {Lin}}, \bibinfo {author} {\bibfnamefont
  {V.}~\bibnamefont {Cirigliano}}, \ and\ \bibinfo {author} {\bibfnamefont
  {T.}~\bibnamefont {Bhattacharya}},\ }\href {\doibase
  10.1103/PhysRevD.98.034503} {\bibfield  {journal} {\bibinfo  {journal} {Phys.
  Rev.}\ }\textbf {\bibinfo {volume} {D98}},\ \bibinfo {pages} {034503}
  (\bibinfo {year} {2018})},\ \Eprint {http://arxiv.org/abs/1806.09006}
  {arXiv:1806.09006 [hep-lat]} \BibitemShut {NoStop}%
\bibitem [{\citenamefont {Follana}\ \emph {et~al.}(2007)\citenamefont {Follana}
  \emph {et~al.}}]{Follana:2006rc}%
  \BibitemOpen
  \bibfield  {author} {\bibinfo {author} {\bibfnamefont {E.}~\bibnamefont
  {Follana}} \emph {et~al.} (\bibinfo {collaboration} {HPQCD Collaboration,
  UKQCD Collaboration}),\ }\href {\doibase 10.1103/PhysRevD.75.054502}
  {\bibfield  {journal} {\bibinfo  {journal} {Phys.Rev.}\ }\textbf {\bibinfo
  {volume} {D75}},\ \bibinfo {pages} {054502} (\bibinfo {year} {2007})},\
  \Eprint {http://arxiv.org/abs/hep-lat/0610092} {arXiv:hep-lat/0610092
  [hep-lat]} \BibitemShut {NoStop}%
\bibitem [{\citenamefont {Bazavov}\ \emph {et~al.}(2013)\citenamefont {Bazavov}
  \emph {et~al.}}]{Bazavov:2012xda}%
  \BibitemOpen
  \bibfield  {author} {\bibinfo {author} {\bibfnamefont {A.}~\bibnamefont
  {Bazavov}} \emph {et~al.} (\bibinfo {collaboration} {MILC Collaboration}),\
  }\href {\doibase 10.1103/PhysRevD.87.054505} {\bibfield  {journal} {\bibinfo
  {journal} {Phys.Rev.}\ }\textbf {\bibinfo {volume} {D87}},\ \bibinfo {pages}
  {054505} (\bibinfo {year} {2013})},\ \Eprint {http://arxiv.org/abs/1212.4768}
  {arXiv:1212.4768 [hep-lat]} \BibitemShut {NoStop}%
\bibitem [{\citenamefont {Bali}\ \emph {et~al.}(2010)\citenamefont {Bali},
  \citenamefont {Collins},\ and\ \citenamefont {Schafer}}]{Bali:2009hu}%
  \BibitemOpen
  \bibfield  {author} {\bibinfo {author} {\bibfnamefont {G.~S.}\ \bibnamefont
  {Bali}}, \bibinfo {author} {\bibfnamefont {S.}~\bibnamefont {Collins}}, \
  and\ \bibinfo {author} {\bibfnamefont {A.}~\bibnamefont {Schafer}},\ }\href
  {\doibase 10.1016/j.cpc.2010.05.008} {\bibfield  {journal} {\bibinfo
  {journal} {Comput.Phys.Commun.}\ }\textbf {\bibinfo {volume} {181}},\
  \bibinfo {pages} {1570} (\bibinfo {year} {2010})},\ \Eprint
  {http://arxiv.org/abs/0910.3970} {arXiv:0910.3970 [hep-lat]} \BibitemShut
  {NoStop}%
\bibitem [{\citenamefont {Blum}\ \emph {et~al.}(2013)\citenamefont {Blum},
  \citenamefont {Izubuchi},\ and\ \citenamefont {Shintani}}]{Blum:2012uh}%
  \BibitemOpen
  \bibfield  {author} {\bibinfo {author} {\bibfnamefont {T.}~\bibnamefont
  {Blum}}, \bibinfo {author} {\bibfnamefont {T.}~\bibnamefont {Izubuchi}}, \
  and\ \bibinfo {author} {\bibfnamefont {E.}~\bibnamefont {Shintani}},\ }\href
  {\doibase 10.1103/PhysRevD.88.094503} {\bibfield  {journal} {\bibinfo
  {journal} {Phys.Rev.}\ }\textbf {\bibinfo {volume} {D88}},\ \bibinfo {pages}
  {094503} (\bibinfo {year} {2013})},\ \Eprint {http://arxiv.org/abs/1208.4349}
  {arXiv:1208.4349 [hep-lat]} \BibitemShut {NoStop}%
\bibitem [{\citenamefont {Sharpe}()}]{Sharpe:2018PQ}%
  \BibitemOpen
  \bibfield  {author} {\bibinfo {author} {\bibfnamefont {S.}~\bibnamefont
  {Sharpe}},\ }\href@noop {} {}\bibinfo {note} {{Private
  Communications}}\BibitemShut {NoStop}%
\bibitem [{\citenamefont {Bhattacharya}\ \emph {et~al.}(2006)\citenamefont
  {Bhattacharya}, \citenamefont {Gupta}, \citenamefont {Lee}, \citenamefont
  {Sharpe},\ and\ \citenamefont {Wu}}]{Bhattacharya:2005rb}%
  \BibitemOpen
  \bibfield  {author} {\bibinfo {author} {\bibfnamefont {T.}~\bibnamefont
  {Bhattacharya}}, \bibinfo {author} {\bibfnamefont {R.}~\bibnamefont {Gupta}},
  \bibinfo {author} {\bibfnamefont {W.}~\bibnamefont {Lee}}, \bibinfo {author}
  {\bibfnamefont {S.~R.}\ \bibnamefont {Sharpe}}, \ and\ \bibinfo {author}
  {\bibfnamefont {J.~M.~S.}\ \bibnamefont {Wu}},\ }\href {\doibase
  10.1103/PhysRevD.73.034504} {\bibfield  {journal} {\bibinfo  {journal} {Phys.
  Rev.}\ }\textbf {\bibinfo {volume} {D73}},\ \bibinfo {pages} {034504}
  (\bibinfo {year} {2006})},\ \Eprint {http://arxiv.org/abs/hep-lat/0511014}
  {arXiv:hep-lat/0511014 [hep-lat]} \BibitemShut {NoStop}%
\bibitem [{\citenamefont {Constantinou}\ \emph {et~al.}(2016)\citenamefont
  {Constantinou}, \citenamefont {Hadjiantonis}, \citenamefont {Panagopoulos},\
  and\ \citenamefont {Spanoudes}}]{Constantinou:2016ieh}%
  \BibitemOpen
  \bibfield  {author} {\bibinfo {author} {\bibfnamefont {M.}~\bibnamefont
  {Constantinou}}, \bibinfo {author} {\bibfnamefont {M.}~\bibnamefont
  {Hadjiantonis}}, \bibinfo {author} {\bibfnamefont {H.}~\bibnamefont
  {Panagopoulos}}, \ and\ \bibinfo {author} {\bibfnamefont {G.}~\bibnamefont
  {Spanoudes}},\ }\href {\doibase 10.1103/PhysRevD.94.114513} {\bibfield
  {journal} {\bibinfo  {journal} {Phys. Rev.}\ }\textbf {\bibinfo {volume}
  {D94}},\ \bibinfo {pages} {114513} (\bibinfo {year} {2016})},\ \Eprint
  {http://arxiv.org/abs/1610.06744} {arXiv:1610.06744 [hep-lat]} \BibitemShut
  {NoStop}%
\bibitem [{\citenamefont {Alexandrou}\ \emph
  {et~al.}(2017{\natexlab{a}})\citenamefont {Alexandrou} \emph
  {et~al.}}]{Alexandrou:2017qyt}%
  \BibitemOpen
  \bibfield  {author} {\bibinfo {author} {\bibfnamefont {C.}~\bibnamefont
  {Alexandrou}} \emph {et~al.},\ }\href {\doibase 10.1103/PhysRevD.96.099906,
  10.1103/PhysRevD.95.114514} {\bibfield  {journal} {\bibinfo  {journal} {Phys.
  Rev.}\ }\textbf {\bibinfo {volume} {D95}},\ \bibinfo {pages} {114514}
  (\bibinfo {year} {2017}{\natexlab{a}})},\ \bibinfo {note} {[Erratum: Phys.
  Rev.D96,no.9,099906(2017)]},\ \Eprint {http://arxiv.org/abs/1703.08788}
  {arXiv:1703.08788 [hep-lat]} \BibitemShut {NoStop}%
\bibitem [{\citenamefont {Alexandrou}\ \emph
  {et~al.}(2017{\natexlab{b}})\citenamefont {Alexandrou}, \citenamefont
  {Constantinou}, \citenamefont {Hadjiyiannakou}, \citenamefont {Jansen},
  \citenamefont {Kallidonis}, \citenamefont {Koutsou}, \citenamefont {Vaquero
  Avilés-Casco},\ and\ \citenamefont {Wiese}}]{Alexandrou:2017oeh}%
  \BibitemOpen
  \bibfield  {author} {\bibinfo {author} {\bibfnamefont {C.}~\bibnamefont
  {Alexandrou}}, \bibinfo {author} {\bibfnamefont {M.}~\bibnamefont
  {Constantinou}}, \bibinfo {author} {\bibfnamefont {K.}~\bibnamefont
  {Hadjiyiannakou}}, \bibinfo {author} {\bibfnamefont {K.}~\bibnamefont
  {Jansen}}, \bibinfo {author} {\bibfnamefont {C.}~\bibnamefont {Kallidonis}},
  \bibinfo {author} {\bibfnamefont {G.}~\bibnamefont {Koutsou}}, \bibinfo
  {author} {\bibfnamefont {A.}~\bibnamefont {Vaquero Avilés-Casco}}, \ and\
  \bibinfo {author} {\bibfnamefont {C.}~\bibnamefont {Wiese}},\ }\href
  {\doibase 10.1103/PhysRevLett.119.142002} {\bibfield  {journal} {\bibinfo
  {journal} {Phys. Rev. Lett.}\ }\textbf {\bibinfo {volume} {119}},\ \bibinfo
  {pages} {142002} (\bibinfo {year} {2017}{\natexlab{b}})},\ \Eprint
  {http://arxiv.org/abs/1706.02973} {arXiv:1706.02973 [hep-lat]} \BibitemShut
  {NoStop}%
\bibitem [{\citenamefont {Green}\ \emph {et~al.}(2017)\citenamefont {Green},
  \citenamefont {Hasan}, \citenamefont {Meinel}, \citenamefont {Engelhardt},
  \citenamefont {Krieg}, \citenamefont {Laeuchli}, \citenamefont {Negele},
  \citenamefont {Orginos}, \citenamefont {Pochinsky},\ and\ \citenamefont
  {Syritsyn}}]{Green:2017keo}%
  \BibitemOpen
  \bibfield  {author} {\bibinfo {author} {\bibfnamefont {J.}~\bibnamefont
  {Green}}, \bibinfo {author} {\bibfnamefont {N.}~\bibnamefont {Hasan}},
  \bibinfo {author} {\bibfnamefont {S.}~\bibnamefont {Meinel}}, \bibinfo
  {author} {\bibfnamefont {M.}~\bibnamefont {Engelhardt}}, \bibinfo {author}
  {\bibfnamefont {S.}~\bibnamefont {Krieg}}, \bibinfo {author} {\bibfnamefont
  {J.}~\bibnamefont {Laeuchli}}, \bibinfo {author} {\bibfnamefont
  {J.}~\bibnamefont {Negele}}, \bibinfo {author} {\bibfnamefont
  {K.}~\bibnamefont {Orginos}}, \bibinfo {author} {\bibfnamefont
  {A.}~\bibnamefont {Pochinsky}}, \ and\ \bibinfo {author} {\bibfnamefont
  {S.}~\bibnamefont {Syritsyn}},\ }\href {\doibase 10.1103/PhysRevD.95.114502}
  {\bibfield  {journal} {\bibinfo  {journal} {Phys. Rev.}\ }\textbf {\bibinfo
  {volume} {D95}},\ \bibinfo {pages} {114502} (\bibinfo {year} {2017})},\
  \Eprint {http://arxiv.org/abs/1703.06703} {arXiv:1703.06703 [hep-lat]}
  \BibitemShut {NoStop}%
\bibitem [{\citenamefont {Bernard}\ \emph {et~al.}(1992)\citenamefont
  {Bernard}, \citenamefont {Kaiser}, \citenamefont {Kambor},\ and\
  \citenamefont {Meissner}}]{Bernard:1992qa}%
  \BibitemOpen
  \bibfield  {author} {\bibinfo {author} {\bibfnamefont {V.}~\bibnamefont
  {Bernard}}, \bibinfo {author} {\bibfnamefont {N.}~\bibnamefont {Kaiser}},
  \bibinfo {author} {\bibfnamefont {J.}~\bibnamefont {Kambor}}, \ and\ \bibinfo
  {author} {\bibfnamefont {U.~G.}\ \bibnamefont {Meissner}},\ }\href {\doibase
  10.1016/0550-3213(92)90615-I} {\bibfield  {journal} {\bibinfo  {journal}
  {Nucl. Phys.}\ }\textbf {\bibinfo {volume} {B388}},\ \bibinfo {pages} {315}
  (\bibinfo {year} {1992})}\BibitemShut {NoStop}%
\bibitem [{\citenamefont {Bernard}\ \emph {et~al.}(1995)\citenamefont
  {Bernard}, \citenamefont {Kaiser},\ and\ \citenamefont
  {Meissner}}]{Bernard:1995dp}%
  \BibitemOpen
  \bibfield  {author} {\bibinfo {author} {\bibfnamefont {V.}~\bibnamefont
  {Bernard}}, \bibinfo {author} {\bibfnamefont {N.}~\bibnamefont {Kaiser}}, \
  and\ \bibinfo {author} {\bibfnamefont {U.-G.}\ \bibnamefont {Meissner}},\
  }\href {\doibase 10.1142/S0218301395000092} {\bibfield  {journal} {\bibinfo
  {journal} {Int. J. Mod. Phys.}\ }\textbf {\bibinfo {volume} {E4}},\ \bibinfo
  {pages} {193} (\bibinfo {year} {1995})},\ \Eprint
  {http://arxiv.org/abs/hep-ph/9501384} {arXiv:hep-ph/9501384 [hep-ph]}
  \BibitemShut {NoStop}%
\bibitem [{\citenamefont {Bernard}\ and\ \citenamefont
  {Meissner}(2007)}]{Bernard:2006gx}%
  \BibitemOpen
  \bibfield  {author} {\bibinfo {author} {\bibfnamefont {V.}~\bibnamefont
  {Bernard}}\ and\ \bibinfo {author} {\bibfnamefont {U.-G.}\ \bibnamefont
  {Meissner}},\ }\href {\doibase 10.1146/annurev.nucl.56.080805.140449}
  {\bibfield  {journal} {\bibinfo  {journal} {Ann. Rev. Nucl. Part. Sci.}\
  }\textbf {\bibinfo {volume} {57}},\ \bibinfo {pages} {33} (\bibinfo {year}
  {2007})},\ \Eprint {http://arxiv.org/abs/hep-ph/0611231}
  {arXiv:hep-ph/0611231 [hep-ph]} \BibitemShut {NoStop}%
\bibitem [{\citenamefont {Bernard}\ and\ \citenamefont
  {Meissner}(2006)}]{Bernard:2006te}%
  \BibitemOpen
  \bibfield  {author} {\bibinfo {author} {\bibfnamefont {V.}~\bibnamefont
  {Bernard}}\ and\ \bibinfo {author} {\bibfnamefont {U.-G.}\ \bibnamefont
  {Meissner}},\ }\href {\doibase 10.1016/j.physletb.2006.06.018} {\bibfield
  {journal} {\bibinfo  {journal} {Phys. Lett.}\ }\textbf {\bibinfo {volume}
  {B639}},\ \bibinfo {pages} {278} (\bibinfo {year} {2006})},\ \Eprint
  {http://arxiv.org/abs/hep-lat/0605010} {arXiv:hep-lat/0605010 [hep-lat]}
  \BibitemShut {NoStop}%
\bibitem [{\citenamefont {Khan}\ \emph {et~al.}(2006)\citenamefont {Khan},
  \citenamefont {Gockeler}, \citenamefont {Hagler}, \citenamefont {Hemmert},
  \citenamefont {Horsley} \emph {et~al.}}]{Khan:2006de}%
  \BibitemOpen
  \bibfield  {author} {\bibinfo {author} {\bibfnamefont {A.~A.}\ \bibnamefont
  {Khan}}, \bibinfo {author} {\bibfnamefont {M.}~\bibnamefont {Gockeler}},
  \bibinfo {author} {\bibfnamefont {P.}~\bibnamefont {Hagler}}, \bibinfo
  {author} {\bibfnamefont {T.}~\bibnamefont {Hemmert}}, \bibinfo {author}
  {\bibfnamefont {R.}~\bibnamefont {Horsley}},  \emph {et~al.},\ }\href
  {\doibase 10.1103/PhysRevD.74.094508} {\bibfield  {journal} {\bibinfo
  {journal} {Phys.Rev.}\ }\textbf {\bibinfo {volume} {D74}},\ \bibinfo {pages}
  {094508} (\bibinfo {year} {2006})},\ \Eprint
  {http://arxiv.org/abs/hep-lat/0603028} {arXiv:hep-lat/0603028 [hep-lat]}
  \BibitemShut {NoStop}%
\bibitem [{\citenamefont {Colangelo}\ \emph {et~al.}(2010)\citenamefont
  {Colangelo}, \citenamefont {Fuhrer},\ and\ \citenamefont
  {Lanz}}]{Colangelo:2010ba}%
  \BibitemOpen
  \bibfield  {author} {\bibinfo {author} {\bibfnamefont {G.}~\bibnamefont
  {Colangelo}}, \bibinfo {author} {\bibfnamefont {A.}~\bibnamefont {Fuhrer}}, \
  and\ \bibinfo {author} {\bibfnamefont {S.}~\bibnamefont {Lanz}},\ }\href
  {\doibase 10.1103/PhysRevD.82.034506} {\bibfield  {journal} {\bibinfo
  {journal} {Phys. Rev.}\ }\textbf {\bibinfo {volume} {D82}},\ \bibinfo {pages}
  {034506} (\bibinfo {year} {2010})},\ \Eprint {http://arxiv.org/abs/1005.1485}
  {arXiv:1005.1485 [hep-lat]} \BibitemShut {NoStop}%
\bibitem [{\citenamefont {de~Vries}\ \emph {et~al.}(2011)\citenamefont
  {de~Vries}, \citenamefont {Timmermans}, \citenamefont {Mereghetti},\ and\
  \citenamefont {van Kolck}}]{deVries:2010ah}%
  \BibitemOpen
  \bibfield  {author} {\bibinfo {author} {\bibfnamefont {J.}~\bibnamefont
  {de~Vries}}, \bibinfo {author} {\bibfnamefont {R.}~\bibnamefont
  {Timmermans}}, \bibinfo {author} {\bibfnamefont {E.}~\bibnamefont
  {Mereghetti}}, \ and\ \bibinfo {author} {\bibfnamefont {U.}~\bibnamefont {van
  Kolck}},\ }\href {\doibase 10.1016/j.physletb.2010.11.042} {\bibfield
  {journal} {\bibinfo  {journal} {Phys.Lett.}\ }\textbf {\bibinfo {volume}
  {B695}},\ \bibinfo {pages} {268} (\bibinfo {year} {2011})},\ \Eprint
  {http://arxiv.org/abs/1006.2304} {arXiv:1006.2304 [hep-ph]} \BibitemShut
  {NoStop}%
\bibitem [{\citenamefont {Akaike}(1974)}]{1100705}%
  \BibitemOpen
  \bibfield  {author} {\bibinfo {author} {\bibfnamefont {H.}~\bibnamefont
  {Akaike}},\ }\href {\doibase 10.1109/TAC.1974.1100705} {\bibfield  {journal}
  {\bibinfo  {journal} {IEEE Transactions on Automatic Control}\ }\textbf
  {\bibinfo {volume} {19}},\ \bibinfo {pages} {716} (\bibinfo {year}
  {1974})}\BibitemShut {NoStop}%
\bibitem [{\citenamefont {Alexandrou}\ \emph {et~al.}(2018)\citenamefont
  {Alexandrou}, \citenamefont {Constantinou}, \citenamefont {Hadjiyiannakou},
  \citenamefont {Jansen}, \citenamefont {Kallidonis}, \citenamefont {Koutsou},\
  and\ \citenamefont {Avilés-Casco}}]{Alexandrou:2018lvq}%
  \BibitemOpen
  \bibfield  {author} {\bibinfo {author} {\bibfnamefont {C.}~\bibnamefont
  {Alexandrou}}, \bibinfo {author} {\bibfnamefont {M.}~\bibnamefont
  {Constantinou}}, \bibinfo {author} {\bibfnamefont {K.}~\bibnamefont
  {Hadjiyiannakou}}, \bibinfo {author} {\bibfnamefont {K.}~\bibnamefont
  {Jansen}}, \bibinfo {author} {\bibfnamefont {C.}~\bibnamefont {Kallidonis}},
  \bibinfo {author} {\bibfnamefont {G.}~\bibnamefont {Koutsou}}, \ and\
  \bibinfo {author} {\bibfnamefont {A.~V.}\ \bibnamefont {Avilés-Casco}},\
  }\bibfield  {booktitle} {\emph {\bibinfo {booktitle} {{Proceedings, 35th
  International Symposium on Lattice Field Theory (Lattice 2017): Granada,
  Spain, June 18-24, 2017}}},\ }\href {\doibase 10.1051/epjconf/201817506003}
  {\bibfield  {journal} {\bibinfo  {journal} {EPJ Web Conf.}\ }\textbf
  {\bibinfo {volume} {175}},\ \bibinfo {pages} {06003} (\bibinfo {year}
  {2018})}\BibitemShut {NoStop}%
\bibitem [{\citenamefont {Nocera}\ \emph {et~al.}(2014)\citenamefont {Nocera},
  \citenamefont {Ball}, \citenamefont {Forte}, \citenamefont {Ridolfi},\ and\
  \citenamefont {Rojo}}]{Nocera:2014gqa}%
  \BibitemOpen
  \bibfield  {author} {\bibinfo {author} {\bibfnamefont {E.~R.}\ \bibnamefont
  {Nocera}}, \bibinfo {author} {\bibfnamefont {R.~D.}\ \bibnamefont {Ball}},
  \bibinfo {author} {\bibfnamefont {S.}~\bibnamefont {Forte}}, \bibinfo
  {author} {\bibfnamefont {G.}~\bibnamefont {Ridolfi}}, \ and\ \bibinfo
  {author} {\bibfnamefont {J.}~\bibnamefont {Rojo}} (\bibinfo {collaboration}
  {NNPDF}),\ }\href {\doibase 10.1016/j.nuclphysb.2014.08.008} {\bibfield
  {journal} {\bibinfo  {journal} {Nucl. Phys.}\ }\textbf {\bibinfo {volume}
  {B887}},\ \bibinfo {pages} {276} (\bibinfo {year} {2014})},\ \Eprint
  {http://arxiv.org/abs/1406.5539} {arXiv:1406.5539 [hep-ph]} \BibitemShut
  {NoStop}%
\bibitem [{\citenamefont {de~Florian}\ \emph {et~al.}(2008)\citenamefont
  {de~Florian}, \citenamefont {Sassot}, \citenamefont {Stratmann},\ and\
  \citenamefont {Vogelsang}}]{deFlorian:2008mr}%
  \BibitemOpen
  \bibfield  {author} {\bibinfo {author} {\bibfnamefont {D.}~\bibnamefont
  {de~Florian}}, \bibinfo {author} {\bibfnamefont {R.}~\bibnamefont {Sassot}},
  \bibinfo {author} {\bibfnamefont {M.}~\bibnamefont {Stratmann}}, \ and\
  \bibinfo {author} {\bibfnamefont {W.}~\bibnamefont {Vogelsang}},\ }\href
  {\doibase 10.1103/PhysRevLett.101.072001} {\bibfield  {journal} {\bibinfo
  {journal} {Phys. Rev. Lett.}\ }\textbf {\bibinfo {volume} {101}},\ \bibinfo
  {pages} {072001} (\bibinfo {year} {2008})},\ \Eprint
  {http://arxiv.org/abs/0804.0422} {arXiv:0804.0422 [hep-ph]} \BibitemShut
  {NoStop}%
\bibitem [{\citenamefont {de~Florian}\ \emph {et~al.}(2009)\citenamefont
  {de~Florian}, \citenamefont {Sassot}, \citenamefont {Stratmann},\ and\
  \citenamefont {Vogelsang}}]{deFlorian:2009vb}%
  \BibitemOpen
  \bibfield  {author} {\bibinfo {author} {\bibfnamefont {D.}~\bibnamefont
  {de~Florian}}, \bibinfo {author} {\bibfnamefont {R.}~\bibnamefont {Sassot}},
  \bibinfo {author} {\bibfnamefont {M.}~\bibnamefont {Stratmann}}, \ and\
  \bibinfo {author} {\bibfnamefont {W.}~\bibnamefont {Vogelsang}},\ }\href
  {\doibase 10.1103/PhysRevD.80.034030} {\bibfield  {journal} {\bibinfo
  {journal} {Phys. Rev.}\ }\textbf {\bibinfo {volume} {D80}},\ \bibinfo {pages}
  {034030} (\bibinfo {year} {2009})},\ \Eprint {http://arxiv.org/abs/0904.3821}
  {arXiv:0904.3821 [hep-ph]} \BibitemShut {NoStop}%
\bibitem [{\citenamefont {Sato}\ \emph {et~al.}(2016)\citenamefont {Sato},
  \citenamefont {Melnitchouk}, \citenamefont {Kuhn}, \citenamefont {Ethier},\
  and\ \citenamefont {Accardi}}]{Sato:2016tuz}%
  \BibitemOpen
  \bibfield  {author} {\bibinfo {author} {\bibfnamefont {N.}~\bibnamefont
  {Sato}}, \bibinfo {author} {\bibfnamefont {W.}~\bibnamefont {Melnitchouk}},
  \bibinfo {author} {\bibfnamefont {S.~E.}\ \bibnamefont {Kuhn}}, \bibinfo
  {author} {\bibfnamefont {J.~J.}\ \bibnamefont {Ethier}}, \ and\ \bibinfo
  {author} {\bibfnamefont {A.}~\bibnamefont {Accardi}} (\bibinfo
  {collaboration} {Jefferson Lab Angular Momentum}),\ }\href {\doibase
  10.1103/PhysRevD.93.074005} {\bibfield  {journal} {\bibinfo  {journal} {Phys.
  Rev.}\ }\textbf {\bibinfo {volume} {D93}},\ \bibinfo {pages} {074005}
  (\bibinfo {year} {2016})},\ \Eprint {http://arxiv.org/abs/1601.07782}
  {arXiv:1601.07782 [hep-ph]} \BibitemShut {NoStop}%
\bibitem [{\citenamefont {Ethier}\ \emph {et~al.}(2017)\citenamefont {Ethier},
  \citenamefont {Sato},\ and\ \citenamefont {Melnitchouk}}]{Ethier:2017zbq}%
  \BibitemOpen
  \bibfield  {author} {\bibinfo {author} {\bibfnamefont {J.~J.}\ \bibnamefont
  {Ethier}}, \bibinfo {author} {\bibfnamefont {N.}~\bibnamefont {Sato}}, \ and\
  \bibinfo {author} {\bibfnamefont {W.}~\bibnamefont {Melnitchouk}},\ }\href
  {\doibase 10.1103/PhysRevLett.119.132001} {\bibfield  {journal} {\bibinfo
  {journal} {Phys. Rev. Lett.}\ }\textbf {\bibinfo {volume} {119}},\ \bibinfo
  {pages} {132001} (\bibinfo {year} {2017})},\ \Eprint
  {http://arxiv.org/abs/1705.05889} {arXiv:1705.05889 [hep-ph]} \BibitemShut
  {NoStop}%
\bibitem [{\citenamefont {Edwards}\ and\ \citenamefont
  {Joo}(2005)}]{Edwards:2004sx}%
  \BibitemOpen
  \bibfield  {author} {\bibinfo {author} {\bibfnamefont {R.~G.}\ \bibnamefont
  {Edwards}}\ and\ \bibinfo {author} {\bibfnamefont {B.}~\bibnamefont {Joo}}
  (\bibinfo {collaboration} {SciDAC Collaboration, LHPC Collaboration, UKQCD
  Collaboration}),\ }\href {\doibase 10.1016/j.nuclphysbps.2004.11.254}
  {\bibfield  {journal} {\bibinfo  {journal} {Nucl.Phys.Proc.Suppl.}\ }\textbf
  {\bibinfo {volume} {140}},\ \bibinfo {pages} {832} (\bibinfo {year}
  {2005})},\ \Eprint {http://arxiv.org/abs/hep-lat/0409003}
  {arXiv:hep-lat/0409003 [hep-lat]} \BibitemShut {NoStop}%
\end{thebibliography}%

\end{document}